\documentclass[a4paper,fleqn]{cas-dc}

\usepackage[numbers]{natbib}

\usepackage{savesym}
\usepackage{nicefrac}
\usepackage{balance}
\usepackage{booktabs}
\usepackage{bigstrut}
\usepackage{rotating}
\usepackage{subfig}
\usepackage[process=auto,crop=pdfcrop]{pstool}
\usepackage{alphalph}

\makeatletter
\providecommand\phantomcaption{\caption@refstepcounter\@captype}
\makeatother

\def\tsc#1{\csdef{#1}{\textsc{\lowercase{#1}}\xspace}}
\tsc{WGM}
\tsc{QE}

   \savesymbol{nomenclature}
   \usepackage{nomencl}
  \restoresymbol{OWN}{nomenclature}
   \RequirePackage{ifthen}
   \renewcommand{\nomgroup}[1]{%
       \ifthenelse{\equal{#1}{G}}{\item[]\item[\textit{Greek Symbols}]}{%
         \ifthenelse{\equal{#1}{S}}{\item[]\item[\\ \textit{Sub/super-scripts}]}{}}}

\newcommand{\etal}{et al. }
\newcommand{\Eq}[1]{Eq. \ref{eq:#1}}
\newcommand{\Fig}[1]{Fig. \ref{fig:#1}}

\newcommand{\Tab}[1]{Tab. \ref{tab:#1}}

\newcommand{\figwidth}{1.\linewidth}
\newcommand{\InsFig}[3]{
  \begin{figure}[tb]
    \centering
    \includegraphics[width=\figwidth, angle=0]{#1}
    \caption{#2}
    \label{fig:#3}
  \end{figure}
}
\newcommand{\InsFigStar}[3]{
  \begin{figure*}[!ht]
    \centering
    \includegraphics[width=\figwidth, angle=0]{#1}
    \caption{#2}
    \label{fig:#3}
  \end{figure*}
}

\begin{document}
\let\WriteBookmarks\relax
\def\floatpagepagefraction{1}
\def\textpagefraction{.001}

\shorttitle{A General Correlation to Predict Solubilities in Binary n-Alkane Mixtures}    

\shortauthors{S. G. Johnsen}  

\title [mode = title]{A Review of Experimental Solubilities and a General Correlation between the Temperature-Dependent Solubility and Solute and Solvent Molar Masses for Binary n-Alkane Mixtures}  



%

\author[1]{Sverre Gullikstad Johnsen}[orcid=0000-0001-5362-8908]

\cormark[1]


\ead{sverre.g.johnsen@sintef.no}



\affiliation[1]{organization={SINTEF},
            city={TRONDHEIM},
            country={NORWAY}}




\begin{abstract}
The solubility of a "heavy" alkane (solute) in a "light" alkane (solvent) is generally temperature dependent.
Moreover, it is determined by the molar masses of the solute and solvent.
In the current paper, published solubility data for binary normal-alkane mixtures is reviewed (solid-liquid equilibrium).
A total of 43 unique solute-solvent data-sets, obtained from a total of 24  published experimental studies, are collected and presented in a systematic manner. 
Based on thermodynamic considerations and the experimental data, it is demonstrated that there is a log-linear relationship between the solubility and the temperature in the dilute range.
Linear regression is employed to 1) obtain data-set-specific solubility-temperature best-fit parameters and 2) obtain a general correlation between the solubility and the solvent and solute molar masses and the temperature.
Finally, it is demonstrated that the developed correlation carries predictive power even for multi-component  mixtures by utilizing solvent and solute average molar masses.
\end{abstract}




\maketitle

\printnomenclature[1.3cm] 
\begin{thenomenclature} 

 \nomgroup{A}

  \item [{$A$,$B$}]\begingroup Linear regression parameters.\nomeqref {0}
		\nompageref{1}
  \item [{$\Delta C_{pm}$}]\begingroup Molar specific heat capacity difference between solid and liquid states of the solute, at the melting point.\nomeqref {0}
		\nompageref{1}
  \item [{$\mathbf e$}]\begingroup Columnn vector of errors.\nomeqref {0}
		\nompageref{1}
  \item [{$f(\mathbf e)$}]\begingroup Best-fit error functional.\nomeqref {0}
		\nompageref{1}
  \item [{$f_1$,$f_2$,$f_3$}]\begingroup Solute- and solvent-dependent solubility parameters.\nomeqref {0}
		\nompageref{1}
  \item [{$\Delta H_m$}]\begingroup Solute molar enthalpy of fusion.\nomeqref {0}
		\nompageref{1}
  \item [{$\Delta H_{tr}$}]\begingroup Solute molar enthalpy of solid-solid transition.\nomeqref {0}
		\nompageref{1}
  \item [{$I$}]\begingroup Number of data-points.\nomeqref {0}
		\nompageref{1}
  \item [{$J$}]\begingroup Number of regression parameters.\nomeqref {0}
		\nompageref{1}
  \item [{$k$}]\begingroup Huber loss function parameter.\nomeqref {0}
		\nompageref{1}
  \item [{$M$}]\begingroup Molar mass.\nomeqref {0}\nompageref{1}
  \item [{$MAD$}]\begingroup Median of absolute deviations.\nomeqref {0}
		\nompageref{1}
  \item [{$N$}]\begingroup Number of carbon atoms in alkane-chain.\nomeqref {0}
		\nompageref{1}
  \item [{$R$}]\begingroup Universal gas constant.\nomeqref {0}
		\nompageref{1}
  \item [{$T$}]\begingroup Temperature.\nomeqref {0}\nompageref{1}
  \item [{$T_m$}]\begingroup Melting temperature\nomeqref {0}
		\nompageref{1}
  \item [{$T_{tr}$}]\begingroup Solid-solid transition temperature.\nomeqref {0}
		\nompageref{1}
  \item [{$\boldsymbol X$}]\begingroup $I\times J$ matrix containing experimental parameters.\nomeqref {0}
		\nompageref{1}
  \item [{$x_s$}]\begingroup Solute solubility (mol-fraction).\nomeqref {0}
		\nompageref{1}
  \item [{$\mathbf Y$}]\begingroup Column vector of calculated  data.\nomeqref {0}
		\nompageref{1}
  \item [{$\mathbf y$}]\begingroup Column vector of experimental data.\nomeqref {0}
		\nompageref{1}

 \nomgroup{G}

  \item [{$\boldsymbol\beta$}]\begingroup Column vector of model parameters.\nomeqref {0}
		\nompageref{1}
  \item [{$\boldsymbol\rho$}]\begingroup Error vector function.\nomeqref {0}
		\nompageref{1}
  \item [{$\gamma$}]\begingroup Activity coefficient.\nomeqref {0}
		\nompageref{1}
  \item [{$\sigma$}]\begingroup Standard deviation.\nomeqref {0}
		\nompageref{1}

 \nomgroup{S}

  \item [{$1$}]\begingroup Solvent.\nomeqref {0}\nompageref{1}
  \item [{$2$}]\begingroup Solute.\nomeqref {0}\nompageref{1}
  \item [{$\hat{}$}]\begingroup Best-fit parameter.\nomeqref {0}
		\nompageref{1}
  \item [{$i$,$j$}]\begingroup Matrix row and column indices.\nomeqref {0}
		\nompageref{1}

\end{thenomenclature}


\section{Introduction}







Petroleum waxes are mainly associated with the aliphatic fraction of crude oil, and normal-alkanes, forming needle-like macro-crystals, are recognized as the main contributor to solid deposits forming during e.g. production and transportation of oil and gas \cite{Misra95}.
Significant costs are associated with the prevention and mitigation of wax precipitation and deposition since accumulation of solid wax in oil pipe-lines may lead to increased operational expenses (e.g. compressors, heating, chemical inhibitors, and man-hours) and reduced production (e.g. diminished flow capacity and periods of shut-in), or in the worst-case abandonment of the entire field \cite{Gluyas2003}.
Being able to predict solid-liquid equilibrium in petroleum systems, is paramount in developing and designing transporting and processing solutions.
Naturally, the oil industry's need for accurate predictions of wax deposition has lead to immense research activity on e.g. thermodynamic modelling and experimental characterisation of waxy oils.
State-of-art modelling allows accurate predictions of phase-behaviour in complex fluids, see e.g. \citet{Coutinho06, Heidariyan19, Shahdi19, Wang20}, and \citet{Ghasemi21}. 

Simplified  model oils are commonly utilized in wax deposition studies and for validation of mathematical models, e.g. by \citet{Singh2000, Paso2004, Wu2004}, and \citet{Johnsen11}.
Although advanced, highly accurate thermodynamic models exist, for wax formation in hydrocarbon systems, the author's own experience as well as that of colleagues show that it is sometimes useful to have a quick and easy way of calculating rough estimates, without the need for extensive input data.
This could be to e.g. evaluate suggested model oils, to initialize computational models, just produce semi-realistic data for student assignments, or whenever it is impractical to perform a rigorous thermodynamic calculation.

In this paper, published solubility data for binary normal-alkane mixtures is reviewed.
A total of 43 unique solute-solvent data-sets, obtained from a total of 24 published experimental studies, are collected and presented in a systematic manner.
It is demonstrated that there is a log-linear relationship between the solubility and the temperature in the dilute range.
Linear regression is employed to 1) obtain data-set-specific solubility-temperature best-fit parameters and 2) obtain a general correlation between the solubility and the solvent and solute molar masses and the temperature.
Finally, it is demonstrated that the developed correlation carries predictive power even for multi-component  mixtures by utilizing solvent and solute average molar masses.

\section{Definitions}
\emph{Normal alkanes}, also known as normal paraffins, are straight carbon-chain molecules saturated with hydrogen atoms such that no branches or double-bonds exist.
The different alkanes are denoted by $CN$, where $N$ indicates the number of carbon atoms in the chain, the \emph{carbon-number}.
It is common to add the prefix \emph{n} to identify straight-chain alkanes.
In this paper, however, the \emph{n} prefix is left out.
The number of hydrogen atoms in the alkane molecule is given by $2\left(N+1\right)$, and the \emph{molar mass} of an alkane is given by 
$
M(N)=\left(14.026N+2.016\right)\nicefrac{g}{mol}
$.
A \emph{binary system} is a mixture consisting of  two alkane species, only.
The \emph{solute} refers to the heavier of the two alkane species, and the \emph{solvent} refers to the lighter. 
The \emph{solubility} (saturation mol-fraction) is defined as the maximum amount of  solute that can be dissolved in a specific quantity of solvent, at a specified temperature.
The current study is limited to solid-liquid phase equilibrium.
A \emph{data-point} is a measured solubility-temperature pair.
The complete set of data-points, from all the literature references, for a specific binary system, will be referred to as a \emph{data-set}.

\emph{Experimental parameters} 
may vary between different experiments, e.g. the system temperature, the solvent or solute properties, or a function of these.
The \emph{model parameters} for a specific model, however,  are constant.
For a set of $I$ experimental solubilities expressed as a column vector, $\mathbf y$, a linear model in $J$ model parameters, can be expressed as
$
  \mathbf Y = \mathbf X\boldsymbol\beta~,
$
where $\mathbf{Y}$ is the column vector of $I$ calculated approximate data, and the experimental and model parameters are represented by the $I\times J$ matrix $\mathbf X$, where $X_{i,1}\equiv1~\forall~ i$, and the column vector $\boldsymbol\beta$~, respectively.
It is required that the number of model parameters exceeds the number of experimental parameters by one.

The aim of \emph{linear regression} is to find a linear model that can be employed to predict or estimate the experimental data with acceptable accuracy.
If a model with only one experimental parameter is chosen,
$
  Y_i=\beta_1+X_{i,2}\beta_2~,
$
and the procedure is referred to as \emph{simple regression}.
If, on the other hand, a model of two or more input parameters is chosen,
$
  Y_i=\beta_1+\sum
_{j=2}^{J\geq3}X_{i,j}\beta_j~,
$
and the procedure is referred to as \emph{multiple regression}.
The \emph{error vector} is defined as
$ \mathbf e=\mathbf{y}-\mathbf{Y}
$
and the \emph{best-fit model parameters} are given by the vector $\boldsymbol{\hat\beta}$, that minimizes the \emph{error functional},
$
  f(\mathbf e)=\sum
  _i^N\rho_i(e_i)~,
$
where each component of the vector function $\boldsymbol\rho$ is a function of the corresponding component of the error vector.
The choice of statistical method determines the representation of the error functional.

\section{Published Solubility Data}

\InsFigStar{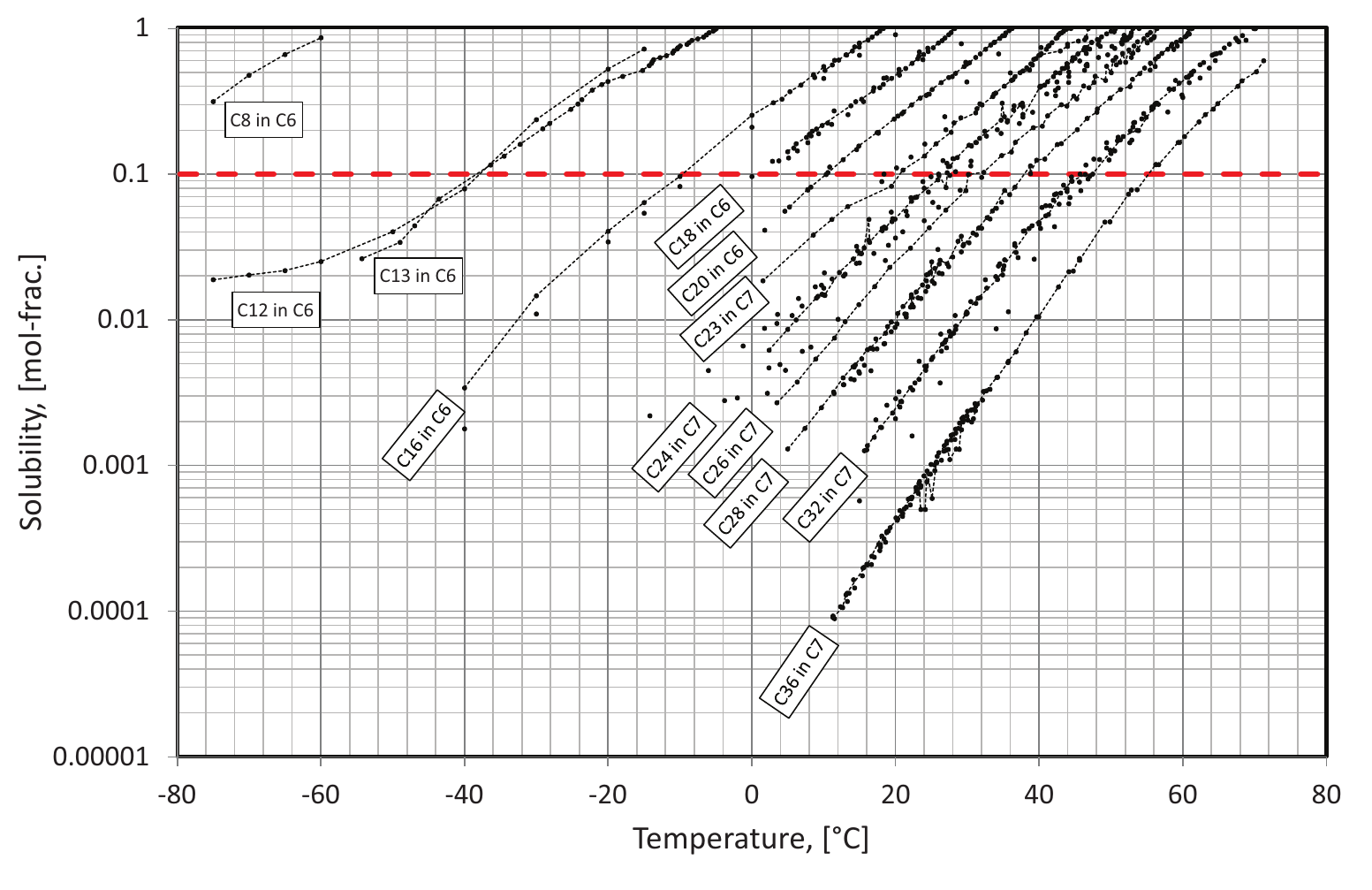}{The complete set of experimental solubility-temperature data. Selected data-sets have been emphasized with labels and black dotted lines, and the 0.1 mol-fraction line is emphasized as the red dashed line.}{SolubilityData}

Experimental solubility data utilized in correlation development in the current paper, have been obtained from 24 publications, for a total of 43 binary systems with solute carbon-numbers ranging from  8 to 36 and solvent carbon-numbers ranging from 3 to 14.
Most data were reported for atmospheric pressure, but propane and butane are gaseous at atmospheric pressure, so the works by \citet{Seyer36} and \citet{Godard37} were performed at the vapour pressure of the solvent.
The solubility data references are summarized in \Tab{SolubilityRef}, and the complete set of experimental data is plotted in \Fig{SolubilityData}.

\citet{Campestrini18} and \citet{Tassin20} provide comprehensive lists of references to published solid-fluid equilibrium data for n-alkanes in methane (C1), ethane (C2), and propane (C3).
These data were not employed for correlation development.

\begin{table*}[!tb]
    \centering
    \caption{Literature references to experimental solubility data for binary alkane mixtures. The references in the hatched table cells did not provide data points below the mole-fraction threshold of $0.1$.}\label{tab:SolubilityRef}%
    \tiny
\begin{tabular}{r|r|r|r|r|r|r|r|r|r|r|r}
\multicolumn{1}{r}{} & \multicolumn{1}{r}{} & \multicolumn{1}{r}{} & \multicolumn{1}{r}{} & \multicolumn{1}{r}{} & \multicolumn{1}{r}{} & \multicolumn{1}{r}{} & \multicolumn{1}{r}{} & \multicolumn{1}{r}{} & \multicolumn{1}{r}{} & \multicolumn{1}{r}{} &  \bigstrut[b]\\
\cline{2-11}\textbf{} & \textbf{Solute\textbackslash Solvent} & \textbf{Propane } & \textbf{Butane } & \textbf{Pentane} & \textbf{Hexane} & \textbf{Heptane } & \textbf{Octane} & \textbf{Decane } & \textbf{Dodecane} & \textbf{Tetradecane } & \textbf{} \bigstrut[t]\\
\textbf{} & \textbf{} & \textbf{(C3)} & \textbf{(C4)} & \textbf{ (C5)} & \textbf{ (C6)} & \textbf{(C7)} & \textbf{ (C8)} & \textbf{(C10)} & \textbf{ (C12)} & \textbf{(C14)} & \textbf{} \bigstrut[b]\\
\cline{2-11}\textbf{} & \textbf{Octane} & \textbf{} & \textbf{} & \textbf{} & \cellcolor{lightgray}\cite{Hoerr51} & \textbf{} & \textbf{} & \textbf{} & \textbf{} & \textbf{} & \textbf{} \bigstrut[t]\\
\textbf{} & \textbf{(C8)} & \textbf{} & \textbf{} & \textbf{} & \cellcolor{lightgray}\textbf{} & \textbf{} & \textbf{} & \textbf{} & \textbf{} & \textbf{} & \textbf{} \bigstrut[b]\\
\cline{2-11}\textbf{} & \textbf{Dodecane} & \textbf{} & \textbf{} & \textbf{} & \cite{Hoerr51} & \textbf{} & \textbf{} & \textbf{} & \textbf{} & \textbf{} & \textbf{} \bigstrut[t]\\
\textbf{} & \textbf{(C12)} & \textbf{} & \textbf{} & \textbf{} & \textbf{} & \textbf{} & \textbf{} & \textbf{} & \textbf{} & \textbf{} & \textbf{} \bigstrut[b]\\
\cline{2-11}\textbf{} & \textbf{Tridecane } &       &       &       & \cite{Morawski05} &       &       &       &       &       & \textbf{} \bigstrut[t]\\
\textbf{} & \textbf{(C13)} &       &       &       &       &       &       &       &       &       & \textbf{} \bigstrut[b]\\
\cline{2-11}\textbf{} & \textbf{Hexadecane} &       &       &       & \cite{Hoerr51}, \cite{Dernini76} & \cite{SelData90} &       &       &       &       & \textbf{} \bigstrut[t]\\
\textbf{} & \textbf{ (C16)} &       &       &       &       &       &       &       &       &       & \textbf{} \bigstrut[b]\\
\cline{2-11}\textbf{} & \textbf{Heptadecane} &       &       &       & \cite{Hoerr51} &       &       &       &       &       & \textbf{} \bigstrut[t]\\
\textbf{} & \textbf{(C17)} &       &       &       &       &       &       &       &       &       & \textbf{} \bigstrut[b]\\
\cline{2-11}\textbf{} & \textbf{Octadecane} &       &       &       & \cellcolor{lightgray}\cite{SelData90} & \cellcolor{lightgray}\cite{Chang83}, \cite{Domanska87} &       & \cellcolor{lightgray}\cite{Sadeghazad00} &       &       & \textbf{} \bigstrut[t]\\
\textbf{} & \textbf{ (C18)} &       &       &       & \cellcolor{lightgray}      &  \cellcolor{lightgray}     &       & \cellcolor{lightgray}      &       &       & \textbf{} \bigstrut[b]\\
\cline{2-11}\textbf{} & \textbf{Nonadecane} &       &       &       &       & \cellcolor{lightgray}\cite{Domanska87} &       &       &       &       & \textbf{} \bigstrut[t]\\
\textbf{} & \textbf{ (C19)} &       &       &       &       & \cellcolor{lightgray}      &       &       &       &       & \textbf{} \bigstrut[b]\\
\cline{2-11}\textbf{} & \textbf{Eicosane} &       &       &       & \cellcolor{lightgray}\cite{SelData90} & \cite{Sadeghazad00}, \cite{SelData87c} &       & \cellcolor{lightgray}\cite{Sadeghazad00} &       &       & \textbf{} \bigstrut[t]\\
\textbf{} & \textbf{ (C20)} &       &       &       & \cellcolor{lightgray}      &       &       &  \cellcolor{lightgray}     &       &       & \textbf{} \bigstrut[b]\\
\cline{2-11}\textbf{} & \textbf{Docosane } &       &       &       & \cellcolor{lightgray}\cite{Kniaz91} & \cellcolor{lightgray}\cite{Floter97} &       &       &       &       & \textbf{} \bigstrut[t]\\
\textbf{} & \textbf{(C22)} &       &       &       & \cellcolor{lightgray}      &  \cellcolor{lightgray}     &       &       &       &       & \textbf{} \bigstrut[b]\\
\cline{2-11}      & \textbf{Tricosane} &       &       &       &  & \cite{Provost98}      &       &       &       &       &  \bigstrut[t]\\
      & \textbf{(C23)} &       &       &       &       &       &       &       &       &       &  \bigstrut[b]\\
\cline{2-11}      & \textbf{Tetracosane} & \cite{Godard37} & \cite{Godard37} & \cite{Godard37} & \cite{Dernini76} & \cite{Brecevic93}, \cite{Roberts94}, \cite{Floter97}, \cite{SelData87f} &       & \cite{Ashbaugh02}, \cite{Johnsen12} & \cite{Brecevic93} &       &  \bigstrut[t]\\
      & \textbf{(C24)} &       &       &       &       &       &       &       &       &       &  \bigstrut[b]\\
\cline{2-11}      & \textbf{Pentacosane} & \textbf{} & \textbf{} &       &       & \cellcolor{lightgray}\cite{Provost98} &       &       &       & \cellcolor{lightgray}\cite{Rakotosaona04} &  \bigstrut[t]\\
      & \textbf{ (C25)} & \textbf{} & \textbf{} &       &       &  \cellcolor{lightgray}     &       &       &       &  \cellcolor{lightgray}     &  \bigstrut[b]\\
\cline{2-11}      & \textbf{Hexacosane} & \textbf{} & \textbf{} &       &       & \cite{SelData87f}, \cite{Provost98} &       &       &       &       &  \bigstrut[t]\\
      & \textbf{ (C26)} & \textbf{} & \textbf{} &       &       &       &       &       &       &       &  \bigstrut[b]\\
\cline{2-11}      & \textbf{Octacosane} & \textbf{} & \textbf{} & \cite{Madsen76} &       & \cite{Madsen76},  \cite{SelData87f}, \cite{Provost98} &       & \cite{Madsen79}, \cite{Ashbaugh02} & \cite{Madsen79} &       &  \bigstrut[t]\\
      & \textbf{ (C28)} & \textbf{} & \textbf{} &       &       &       &       &       &       &       &  \bigstrut[b]\\
\cline{2-11}      & \textbf{Dotriacontane} & \cite{Seyer36} & \cite{Seyer36} & \cite{Madsen76} & \cite{Seyer38},\cite{Hoerr51} & \cite{Hildebrand49}, \cite{Madsen76}, \cite{Chang83}, \cite{Roberts94} & \cite{Seyer38} & \cite{Seyer38},   \cite{Ashbaugh02} & \cite{Seyer38} &       &  \bigstrut[t]\\
      & \textbf{ (C32)} &       &       &       &       &       &       &       &       &       &  \bigstrut[b]\\
\cline{2-11}      & \textbf{Hexatriacontane} & \textbf{} & \textbf{} &       & \cite{Madsen79} & \cite{Madsen76}, \cite{Madsen79}, \cite{Roberts94},\cite{Jennings05} & \cite{Madsen79} & \cite{Madsen79}, \cite{Ashbaugh02} & \cite{Madsen79} &       &  \bigstrut[t]\\
      & \textbf{ (C36)} & \textbf{} &       &       &       &       &       &       &       &       &  \bigstrut[b]\\
\cline{2-11}\multicolumn{1}{r}{} & \multicolumn{1}{r}{} & \multicolumn{1}{r}{\textbf{}} & \multicolumn{1}{r}{} & \multicolumn{1}{r}{} & \multicolumn{1}{r}{} & \multicolumn{1}{r}{} & \multicolumn{1}{r}{} & \multicolumn{1}{r}{} & \multicolumn{1}{r}{} & \multicolumn{1}{r}{} &  \bigstrut[t]\\
\end{tabular}%

    \normalsize
\end{table*}

\Fig{SolubilityData} suggests that solubilities can be approximated by a log-linear relationship with the system temperature in the dilute range.
The deviation from this behaviour increases for increasing solubility, and somewhat arbitrarily it was decided to focus on data below a mol-fraction of $0.1$.
This disqualifies, from the current study, eleven data-sets due to lack of data-points below the limit (C8-C6, C18-C6, C18-C7, C18-C10, C19-C7, C20-C6, C20-C10, C22-C6, C22-C7, C25-C7, C25-C14).

\section{Theory and Method}\label{sec:theory}
The solubility of solids in liquids, in terms of the mol-fraction, can be modelled by  \cite{Weimer65,Choi83}
\begin{equation}\label{eq:ModSolEq}
  \ln x_s=\ln f_1+f_2\ln T+\frac{f_3}{T}~,
\end{equation}
where
$f_1 \equiv f_1(M_{1},M_{2}) = \frac{T_m^{-f_2}}{\gamma_2}\exp\left[f_2+\nicefrac{\left(\frac{\Delta H_m}{T_m}+\sum\limits_{tr}\frac{\Delta H_{tr}}{T_{tr}}\right)}{R}\right]$, 
$f_2 \equiv f_2(M_{2}) = -\nicefrac{\Delta C_{pm}}{R}$, and 
$f_3 \equiv f_3(M_{2}) = \left[f_2 T_m-\right.$ $\left.\nicefrac{\left(\Delta H_m + \sum\limits_{tr}\Delta H_{tr}\right)}{R}\right]$,
 $M_1$ and $M_2$ are the solvent and solute molar masses, respectively, $\gamma$ is the activity coefficient, $\Delta H_m$  is the molar solute enthalpy of fusion, $R$ is the universal gas constant, $T$ is the system temperature,
 $T_m$ is the solute melting-point temperature, $\Delta C_{pm}$ is the molar heat capacity difference between solid and liquid states of the solute at $T_{m}$, and $\Delta H_{tr}$ and $T_{tr}$ are the molar enthalpy and temperature of solid-solid phase transitions, respectively.

The 1st order Taylor expansion of \Eq{ModSolEq}, around some temperature $T^*<T_m$, is
\begin{equation}\label{eq:SimplifiedSolEq}
 \ln x_s\approx A+B\cdot T~,
\end{equation}
where
$A\equiv A(M_{1},M_{2},T^*)=\ln f_1+f_2\ln T^*+\nicefrac{f_3}{T^*}-BT^*$, and $B\equiv B(M_{2}, T^*) = \nicefrac{f_2}{T^*}-\nicefrac{f_3}{{T^*}^2}$.
Due to the observed log-linearity of the solubility, below the mol-fraction of $0.1$, $A$ and $B$ are expected to be independent of $T^*$.
It has been assumed that the activity coefficient, melting and transition temperatures, heat capacities and enthalpies can be expressed as functions of the molar masses, only.
That is, effects of e.g. pressure or molecular structure/nature have not been considered.

Common methods of linear regression include Least-squares regression (LS) and Least absolute deviations regression (LAD).
The current paper employs the Huber loss function \cite{Huber64, Birkes93}, which combines the strengths of the LAD (robustness) and the LS (accuracy) methods. 
For the Huber method, the vector function $\boldsymbol\rho$  may be expressed as 
\begin{equation}
  \rho_i(e_i)=
    \begin{cases}
      e_i^2 & \mbox{if } -k\leq e_i\leq k\\
      2k|e_i|-k^2 & \mbox{if } e<-k \mbox{ or } k<e
    \end{cases}
\end{equation}
where, $k=1.5\sigma$.
$\sigma$ is an estimate of the standard deviation of the population random errors, and for normally distributed errors, $\sigma=1.483MAD$ gives a good estimate. 
$MAD$ is the median of the absolute deviations, $|e_i|$.
This is a robust method that performs reasonably well even when the basic assumptions of the statistics are false \cite{Birkes93}. 

\section{Results}\label{sec:results}
\begin{figure}[tb]
  \centering
  \subfloat[$\hat A$ vs. $M_{1}$\label{fig:AvsMw1}]{\includegraphics[width=\figwidth]{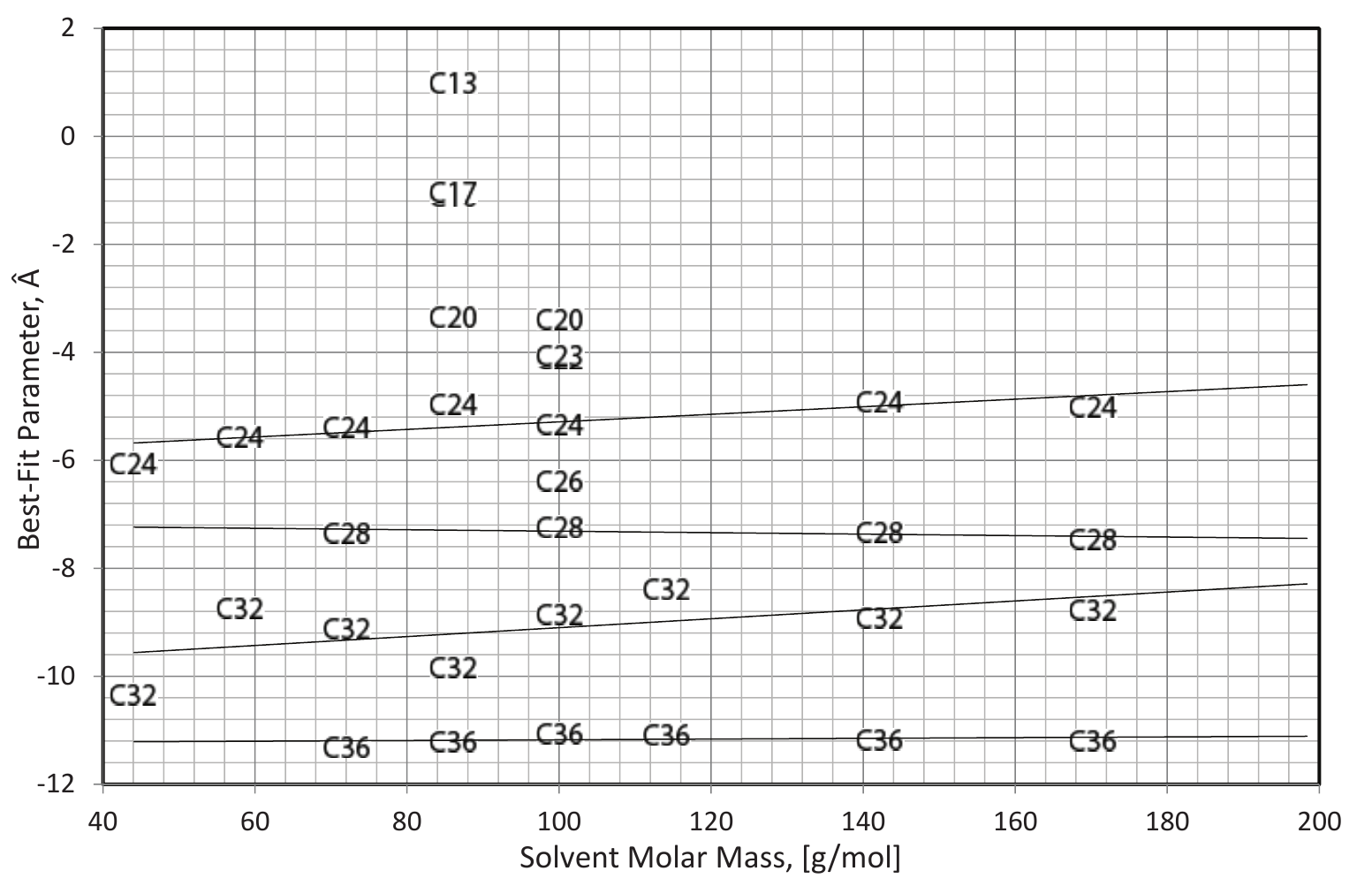}}\\
  \subfloat[$\hat A$ vs. $M_{2}$]{\includegraphics[width=\figwidth]{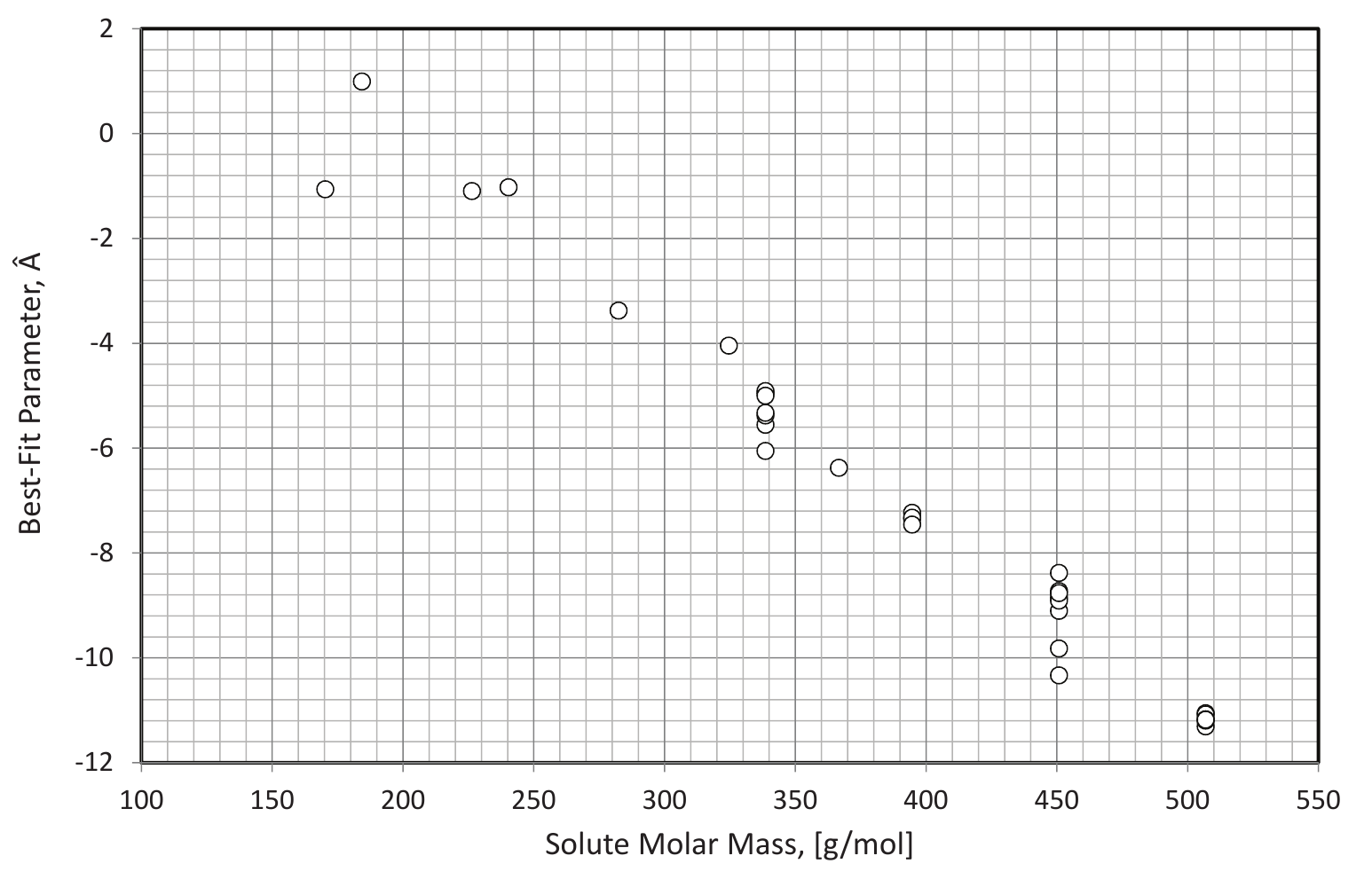}}
  \caption{Data-set specific best-fit parameter $\hat A$ relation to the solvent and solute  molar masses.}
  \label{fig:AvsMw}
\end{figure}

\begin{figure}[tb]
  \centering
  \subfloat[$\hat B$ vs. $M_{1}$\label{fig:BvsMw1}]{\includegraphics[width=\figwidth]{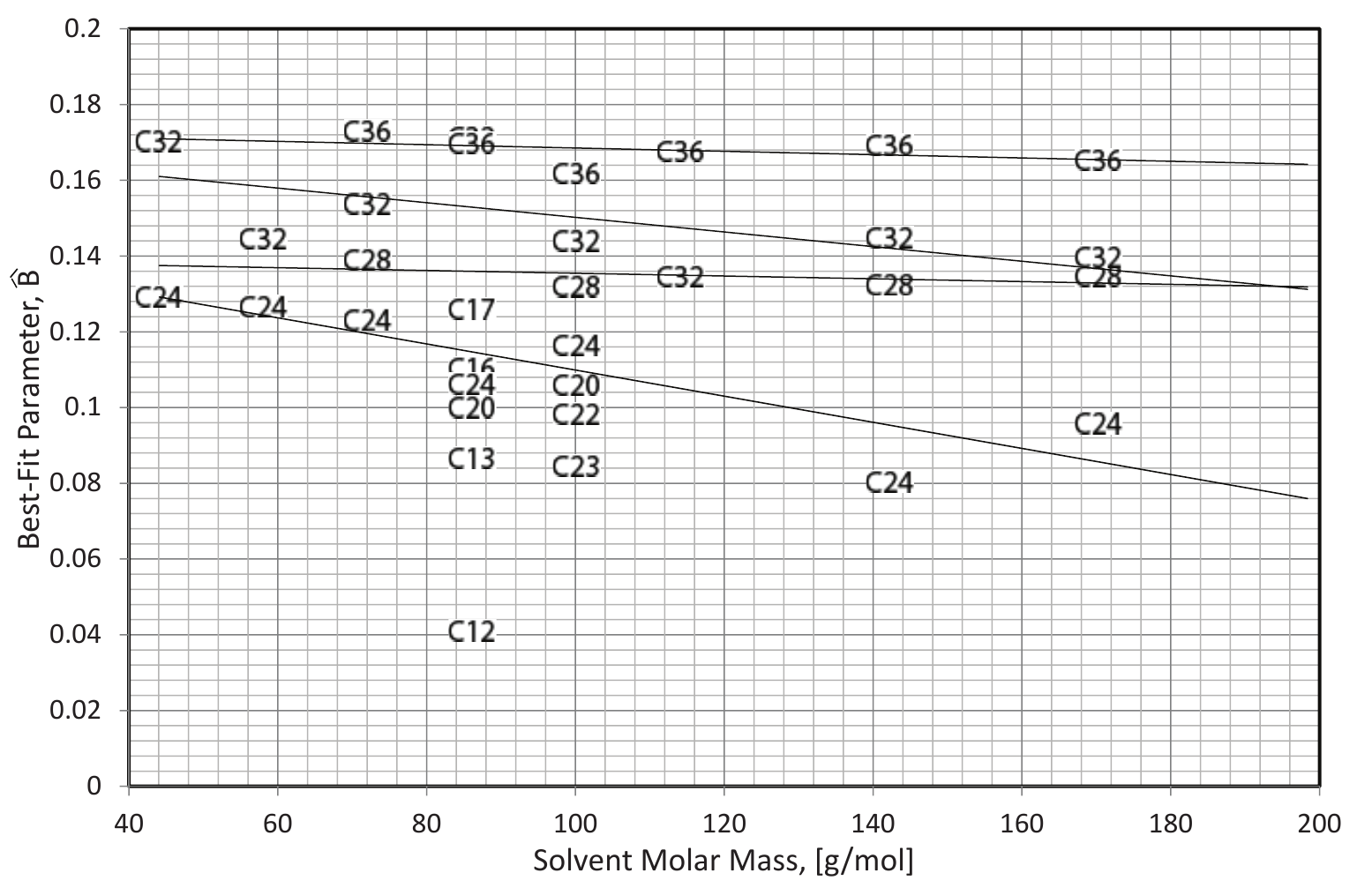}}\\
  \subfloat[$\hat B$ vs. $M_{2}$\label{fig:BvsMw2}]{\includegraphics[width=\figwidth]{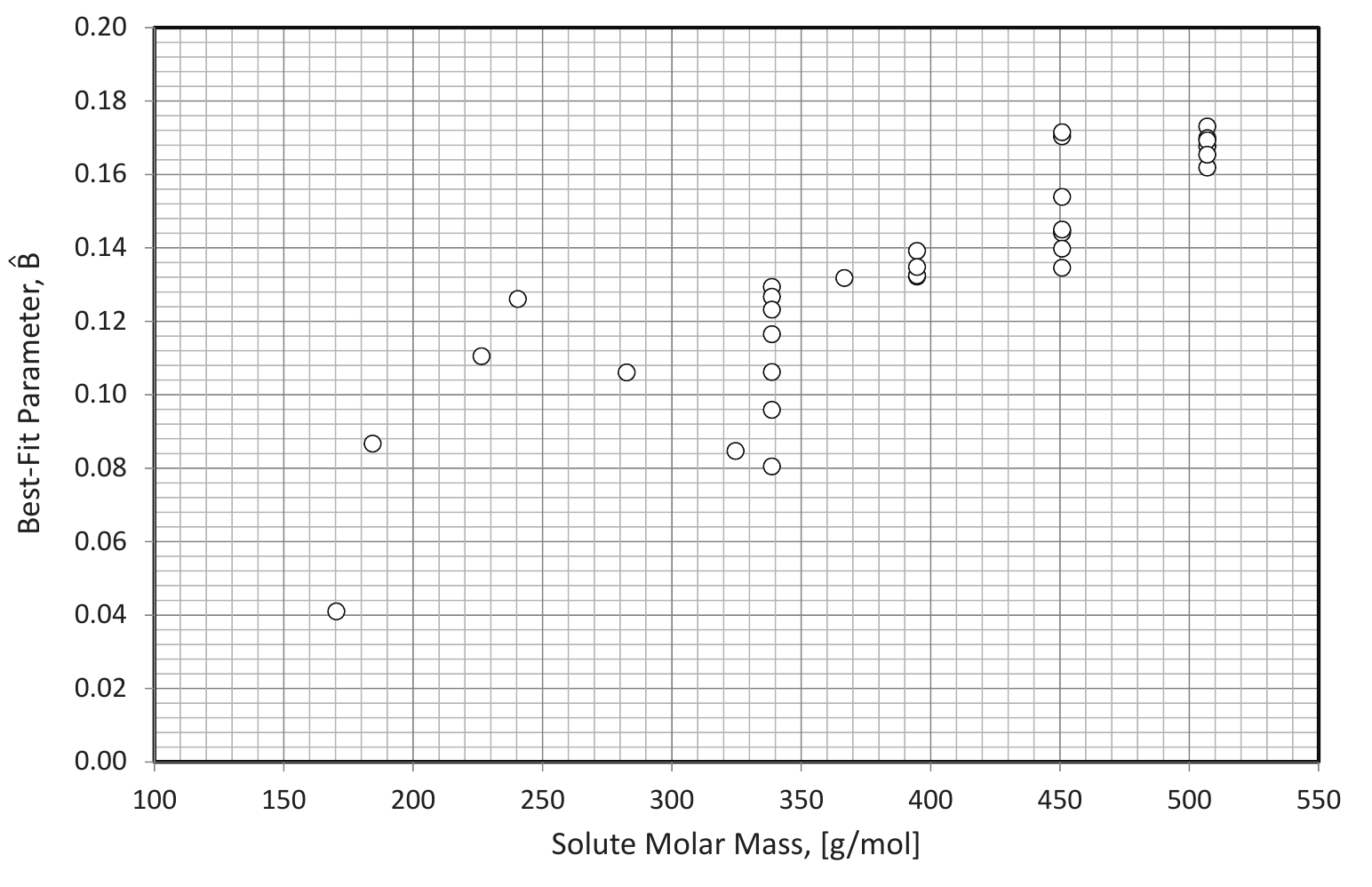}}
  \caption{Data-set specific best-fit parameter $\hat B$ relation to the solvent and solute  molar masses.}
  \label{fig:BvsMw}
\end{figure}

To investigate the $A$ and $B$ (\Eq{SimplifiedSolEq}) dependency on the solvent and solute molar masses, simple regression was performed on each data-set to obtain data-set-specific best-fit parameters.
These are plotted against the molar masses in figures \ref{fig:AvsMw} and \ref{fig:BvsMw} and are cited in \Tab{AB}.
In figures \ref{fig:AvsMw1} and \ref{fig:BvsMw1} linear trends are added for the four series with more than two best-fit points (C24, C28, C32 and C36).
It is clear that there are linear relationships between $A$ and $B$ and the molar masses.
Although some of the $A$ and $B$ outliers seem to deviate significantly from the linear trends, no data-sets were disqualified for this reason.
Some of the scatter can be explained by the dependency on the other molar mass, and in fact, the final regression expression was not very sensitive to elimination of the outliers.
This owes to the robustness of the Huber method.

\begin{table*}[!tb]
    \centering
    \caption{Data-set specific solubility best-fit parameters, $A$ and $B$ (\Eq{SimplifiedSolEq}), obtained for solubilities below a mol-fraction of $0.1$.}\label{tab:AB}%
    \tiny
\begin{tabular}{r|r|r|r|r|r|r|r|r|r|r|r}
\multicolumn{1}{r}{} & \multicolumn{1}{r}{} & \multicolumn{1}{r}{} & \multicolumn{1}{r}{} & \multicolumn{1}{r}{} & \multicolumn{1}{r}{} & \multicolumn{1}{r}{} & \multicolumn{1}{r}{} & \multicolumn{1}{r}{} & \multicolumn{1}{r}{} & \multicolumn{1}{r}{} &  \bigstrut[b]\\
\cline{2-11}      & \textbf{Solute\textbackslash Solvent} & \textbf{Propane } & \textbf{Butane } & \textbf{Pentane} & \textbf{Hexane} & \textbf{Heptane } & \textbf{Octane} & \textbf{Decane } & \textbf{Dodecane} & \textbf{Tetradecane } &  \bigstrut[t]\\
      & \textbf{} & \textbf{(C3)} & \textbf{(C4)} & \textbf{ (C5)} & \textbf{ (C6)} & \textbf{(C7)} & \textbf{ (C8)} & \textbf{(C10)} & \textbf{ (C12)} & \textbf{(C14)} &  \bigstrut[b]\\
\cline{2-11}      & \textbf{Octane} &       &       &       &       &       &       &       &       &       &  \bigstrut[t]\\
      & \textbf{(C8)} &       &       &       &       &       &       &       &       &       &  \bigstrut[b]\\
\cline{2-11}      & \textbf{Dodecane} &       &       &       & -1.062 &       &       &       &       &       &  \bigstrut[t]\\
      & \textbf{(C12)} &       &       &       & 0.0410 &       &       &       &       &       &  \bigstrut[b]\\
\cline{2-11}      & \textbf{Tridecane } &       &       &       & 0.993 &       &       &       &       &       &  \bigstrut[t]\\
      & \textbf{(C13)} &       &       &       & 0.0867 &       &       &       &       &       &  \bigstrut[b]\\
\cline{2-11}      & \textbf{Hexadecane} &       &       &       & -1.100 &       &       &       &       &       &  \bigstrut[t]\\
      & \textbf{ (C16)} &       &       &       & 0.1105 &       &       &       &       &       &  \bigstrut[b]\\
\cline{2-11}      & \textbf{Heptadecane} &       &       &       & -1.025 &       &       &       &       &       &  \bigstrut[t]\\
      & \textbf{(C17)} &       &       &       & 0.1261 &       &       &       &       &       &  \bigstrut[b]\\
\cline{2-11}      & \textbf{Octadecane} &       &       &       &       &       &       &       &       &       &  \bigstrut[t]\\
      & \textbf{ (C18)} &       &       &       &       &       &       &       &       &       &  \bigstrut[b]\\
\cline{2-11}      & \textbf{Nonadecane} &       &       &       &       &       &       &       &       &       &  \bigstrut[t]\\
      & \textbf{ (C19)} &       &       &       &       &       &       &       &       &       &  \bigstrut[b]\\
\cline{2-11}      & \textbf{Eicosane} &       &       &       & -3.340 & -3.379 &       &       &       &       &  \bigstrut[t]\\
      & \textbf{ (C20)} &       &       &       & 0.100 & 0.1061 &       &       &       &       &  \bigstrut[b]\\
\cline{2-11}      & \textbf{Docosane } &       &       &       &       & -4.112 &       &       &       &       &  \bigstrut[t]\\
      & \textbf{(C22)} &       &       &       &       & 0.098 &       &       &       &       &  \bigstrut[b]\\
\cline{2-11}      & \textbf{Tricosane} &       &       &       &       & -4.048 &       &       &       &       &  \bigstrut[t]\\
      & \textbf{(C23)} &       &       &       &       & 0.0847 &       &       &       &       &  \bigstrut[b]\\
\cline{2-11}      & \textbf{Tetracosane} & -6.055 & -5.554 & -5.380 & -4.955 & -5.327 &       & -4.913 & -5.002 &       &  \bigstrut[t]\\
      & \textbf{(C24)} & 0.1294 & 0.1267 & 0.1232 & 0.1062 & 0.1165 &       & 0.0805 & 0.0959 &       &  \bigstrut[b]\\
\cline{2-11}      & \textbf{Pentacosane} &       &       &       &       &       &       &       &       &       &  \bigstrut[t]\\
      & \textbf{ (C25)} &       &       &       &       &       &       &       &       &       &  \bigstrut[b]\\
\cline{2-11}      & \textbf{Hexacosane} &       &       &       &       & -6.375 &       &       &       &       &  \bigstrut[t]\\
      & \textbf{ (C26)} &       &       &       &       & 0.1318 &       &       &       &       &  \bigstrut[b]\\
\cline{2-11}      & \textbf{Octacosane} &       &       & -7.345 &       & -7.233 &       & -7.329 & -7.459 &       &  \bigstrut[t]\\
      & \textbf{ (C28)} &       &       & 0.1392 &       & 0.1322 &       & 0.1325 & 0.1348 &       &  \bigstrut[b]\\
\cline{2-11}      & \textbf{Dotriacontane} & -10.335 & -8.728 & -9.106 & -9.824 & -8.852 & -8.381 & -8.910 & -8.767 &       &  \bigstrut[t]\\
      & \textbf{ (C32)} & 0.1704 & 0.1447 & 0.1539 & 0.1715 & 0.1441 & 0.1346 & 0.1450 & 0.1398 &       &  \bigstrut[b]\\
\cline{2-11}      & \textbf{Hexatriacontane} &       &       & -11.307 & -11.200 & -11.060 & -11.081 & -11.171 & -11.182 &       &  \bigstrut[t]\\
      & \textbf{ (C36)} &       &       & 0.1731 & 0.1699 & 0.1619 & 0.1677 & 0.1693 & 0.1654 &       &  \bigstrut[b]\\
\cline{2-11}\multicolumn{1}{r}{} & \multicolumn{1}{r}{} & \multicolumn{1}{r}{} & \multicolumn{1}{r}{} & \multicolumn{1}{r}{} & \multicolumn{1}{r}{} & \multicolumn{1}{r}{} & \multicolumn{1}{r}{} & \multicolumn{1}{r}{} & \multicolumn{1}{r}{} & \multicolumn{1}{r}{} &  \bigstrut[t]\\
\end{tabular}%

    \normalsize
\end{table*}

Having revealed a linear relationship between the regression parameters, $A$ and $B$, and the molar masses, multiple regression was performed on the entire set of data-points, from all the data-sets, to obtain the best-fit model 
\begin{multline}\label{eq:GenCorr}
  \ln{x_{s}}\approx\left(6.435-6.627\cdot10^{-4}M_{1}-3.446\cdot10^{-2}M_{2}\right)\\
          +\left(1.499-2.989\cdot10^{-2}M_{2}\right)\nicefrac{T}{100}~,
\end{multline}
where the temperature is in $^\circ C$, and any $M_1$ dependency of $B$ is neglected in accordance with the discussion in Section \ref{sec:theory} (\Eq{SimplifiedSolEq}). 

In the appendix (\Fig{SolubilityCurves}), all the solubility data-sets are presented.
In addition, data-set-specific best-fit trend-lines (\Eq{SimplifiedSolEq})(red line) and the general correlation (\Eq{GenCorr})(blue line) are shown.
The \Eq{GenCorr} predictions alone, are drawn for the data-sets with no data-points below the $0.1$ mol-fraction.

\section{Discussion}
Several authors, e.g. \citet{Jennings05}, have pointed out that solubility data is correlated with the solute melting temperature, so that the solubility curves collapse onto each other if plotted against $(T-T_m)$.
The end-point of the solubility curve $x_{s}$, at pure solute, is of course at the solute melting point.
Thus, all solubility curves should terminate in the same point, in an $x_{s}$ vs. $(T-T_{m})$ plot.
The starting point of the curve, however, at pure solvent, will be at the melting  point of the solvent, which varies depending on the solvent molar mass.
The melting temperature of n-alkanes ($^\circ C$) can be approximated by e.g. the Dollhopf correlation \citep{Dollhopf81, Laux99},
\begin{equation}
  \label{eq:Dollhopf}
  T_{m}(N)=\frac{414.6}{1+\nicefrac{6.86}{N}}-273.15~,
\end{equation}
where $N$ is the number of carbon atoms in the alkane-chain.
In \Fig{SolubilityDataTm}, the complete set of reviewed solubility data are plotted against $(T-T_{m})$, using melting temperatures obtained from \Eq{Dollhopf}.
It can indeed be seen that most of the solubility curves gather in a narrow band.
The C13 in C6 data-set (\citet{Morawski05}), identified by red circles in \Fig{SolubilityDataTm}, stands out, however.
It is noted that while \Eq{Dollhopf} produces $T_{m,C13}=-1.76^\circ C$, \citet{Morawski05} reported a C13 melting temperature of $-5.55^\circ C$.
Using the latter improves the C13 in C6 solubility curve by shifting the $x_{s}=1.0$ point to $(T-T_{m})\approx0$ (see black circles in \Fig{SolubilityDataTm}), but the general impression is that the measured C13 in C6 solubilities might be unnaturally high.
\Fig{C13C6} illustrates how the general correlation (\Eq{GenCorr}) generally under-predicts the C13 in C6 data.

It is observed that melting temperatures are prone to errors depending on the method of measurement and the purity of the substance; e.g. isomerization may affect the melting temperature significantly. 
Presuming that the melting temperature is chiefly depending on the molar mass, the solubility's dependency on the solute melting temperature is warranted by the molar mass dependency of the $A$ and the $B$ in \Eq{SimplifiedSolEq}. 
Thus, the current paper did not take the solute melting temperature as input for the developed correlation.

\InsFig{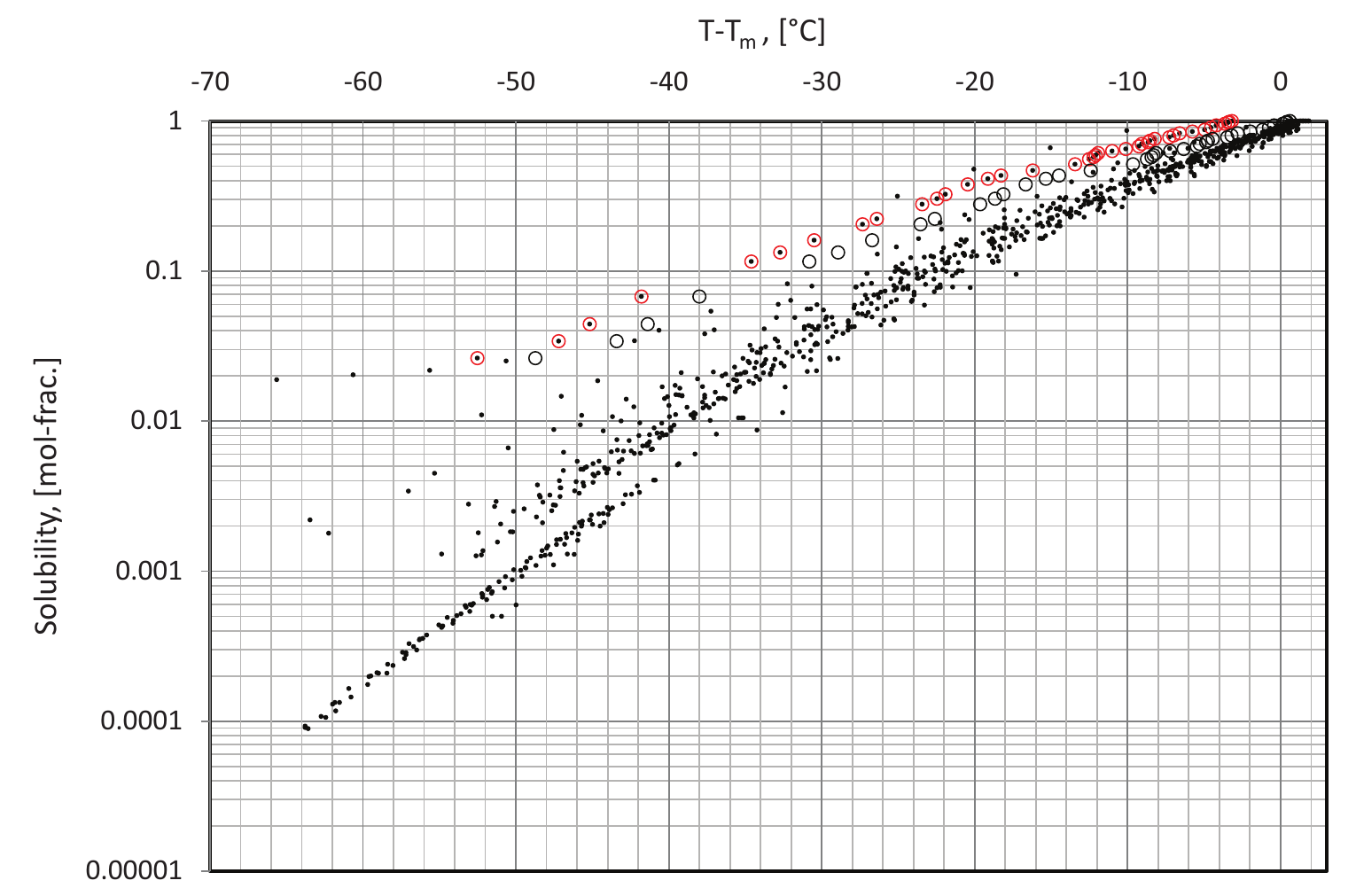}{\emph{Black dots}: The complete set of experimental solubility data plotted against $(T-T_{m})$, using \citet{Dollhopf81} melting temperatures (\Eq{Dollhopf}). \emph{Red circles}: The \citet{Morawski05} C13 in C6 data-set. \emph{Black circles}: The Morawski \etal data-set using their own wax melting temperature instead of the Dollhopf \etal melting temperature.}{SolubilityDataTm}

Several of the referenced authors fail to state the purity or nature of the solutes and solvents used in their studies.
It is suspected, however, that impurities only introduce minor errors in the experimental solubility data. 
\citet{Provost98} stated: ``It is shown that the nature of the solvent has no major influence on the solubility...'', and \citet{Rakotosaona04} concluded that the solubility of a multi-component wax is similar to that of the single-component wax whose carbon-number is equal to the mixture average carbon-number. 
These statements indicate that \Eq{GenCorr} may be utilized or adapted to more complex systems than were studied in this paper.
This conjecture is supported by the good agreement between \Eq{GenCorr} and the experimental solubilities for paraffin waxes in petroleum distillates obtained by \citet{Berne-Allen38}, as demonstrated in \Fig{BerneAllenComp}.
Berne-Allen and Work did not elaborate on the purity of the waxes employed in their study, but specified: ``the solvents were selected with the point in view of obtaining a wide spread over all the lighter fractions from petroleum.''
The properties of the Berne-Allen-Work Solvent 1 and waxes 1-5 are cited in \Tab{BerneAllen}.

\begin{table*}[!tb]
    \centering
    \caption{Properties of solvent and waxes employed by \citet{Berne-Allen38}.}\label{tab:BerneAllen}%
    \footnotesize
\begin{tabular}{r|rr|rrr|r}
\multicolumn{1}{r}{} &       & \multicolumn{1}{r}{} &       &       & \multicolumn{1}{r}{} &  \bigstrut[b]\\
\cline{2-6}      &       &       & B.Pt. /M.Pt., $[^\circ C]$ & Molar mass, $[\nicefrac{g}{mol}]$ & Sp.grav. &  \bigstrut\\
\cline{2-6}      & Solvent & 1     & 105   & 105   & 0.722 &  \bigstrut\\
\cline{2-6}      & Wax   & 1     & 49.9  & 333   &       &  \bigstrut[t]\\
      &       & 2     & 52.8  & 346   &       &  \\
      &       & 3     & 55.6  & 356   &       &  \\
      &       & 4     & 60.3  & 380   &       &  \\
      &       & 5     & 64.4  & 408   &       &  \bigstrut[b]\\
\cline{2-6}\multicolumn{1}{r}{} &       & \multicolumn{1}{r}{} &       &       & \multicolumn{1}{r}{} &  \bigstrut[t]\\
\end{tabular}%

    \normalsize
\end{table*}

\InsFig{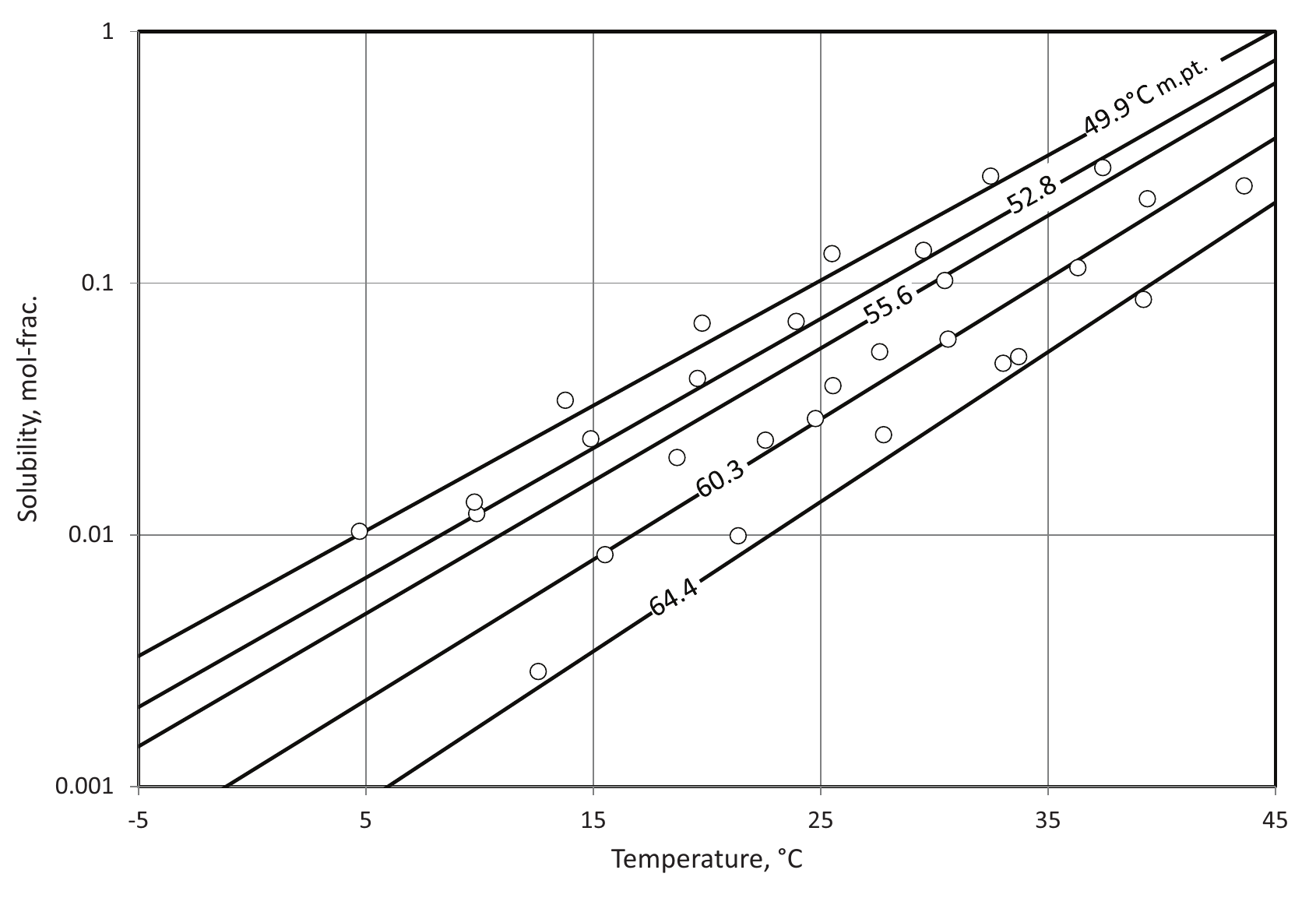}{Comparison of calculated solubilities (\Eq{GenCorr}) with experimental solubilities for petroleum waxes in a multi-component petroleum distillate  (Solvent 1)\cite{Berne-Allen38}. Waxes are identified by their melting points, and molar masses, as input for \Eq{GenCorr}, were taken from \cite{Berne-Allen38}.}{BerneAllenComp}

In the referenced literature, various methods of obtaining the solubilities were exercised.
The two main strategies were 1) reducing the temperature and 2) heating the sample, looking for the first crystal to precipitate out or the last crystal to melt, respectively.
\citet{Ashbaugh02,Johnsen12,Sadeghazad00,Seyer36} and \citet{Seyer38} stated explicitly that the solubilities were found by observing the first crystals precipitate out.
\citet{Dernini76} and \citet{Madsen76,Madsen79} did not clearly state what method they employed.
The remaining authors established the saturation point by observing the last crystals dissolve.
\citet{Dernini76}, \citet{Seyer36} and \citet{Seyer38} stated that there was good agreement between saturation temperatures obtained by heating and cooling (less than $0.1^\circ C$ difference).
Good agreement may not always be the case, however.
\citet{Seyer38} commented that there typically may be a significant difference between the dissolution and precipitation temperatures recorded for heavier alkanes. 
In the case of super-saturation, the solubility may be severely over-predicted.
No obvious signs of super-saturation were identified in the experimental data, but it may be suspected that an effect of super-saturation is to reduce the data-set-specific solubility slope, $B$.
It is observed, in \Fig{C36C10}, that the \citet{Ashbaugh02} data ascends more slowly than the \citet{Madsen79} data.
Furthermore, the C24-C10 data-set, comprised from the \citet{Ashbaugh02} and \citet{Johnsen12} data, produced one of the most severe outliers in \Fig{BvsMw2} although there is good agreement between the two experiments.
The C32-C10 data-set, comprised from the \citet{Ashbaugh02} and \citet{Seyer38} data, did not give evidence of such an effect, however.
 
The current regression analysis was based on a log-linear relation between solubility and temperature, but the deviation from such behaviour increases for increasing solubility as seen in figs. \ref{fig:SolubilityData} and \ref{fig:SolubilityCurves}.
Therefore, an upper validity limit is required, for the regression analysis.
The analysis was thus performed on the subset of the referenced experimental data, with reported solubilities below $0.1$ (dilute solution).
This upper solubility limit was chosen somewhat arbitrarily with the aim of including as much data as possible at the same time as getting as good a curve fit as possible.
On one hand, for a higher limit, more data would be included in the regression analysis, but the increasing deviation from log-linear behaviour would potentially cause a generally less accurate curve-fit.
On the other hand, for a lower limit, less data would be included, which might also result in a less accurate curve fit.
Consequently, it was accepted that the resulting curve fit would be inaccurate above the upper solubility limit.
It is observed, however, that for some data series, the general correlation (\Eq{GenCorr}) reproduces data  above the the upper solubility limit accurately (see e.g. figs. \ref{fig:C16C6}(C16 in C6), \ref{fig:C18C6}-\ref{fig:C18C10}(C18 in C6, C7, and C10), and \ref{fig:C20C6}(C20 in C6), in the appendix).
Above the upper solubility limit, the tendency is that the general correlation under/over-predicts the solubility for solute carbon-numbers below/above $18$, respectively.
The possibility to find a more advanced correlation that will reproduce the experimental data more accurately, for a wider range of temperatures, has not been ruled out by the current study.

The current study used mol-fractions in presenting data and establishing the general correlation in \Eq{GenCorr}.
It is noted that the presentation would look somewhat different if e.g. mass-fractions were used,
due to the non-linear relationship between mol- and mass-fractions and the large span in solute-solvent molar mass ratios considered.
Although mol-fractions were preferred in the present study, the conclusions would be similar if mass-fractions were used instead.
 
Performing the best-fit procedure in the manner described in Section \ref{sec:results} means that all data-points are given the same weight.
Furthermore, this means that the procedure gives more total weight to the data-sets with many data-points than to the data-sets with fewer data-points.
Since a majority of the referenced data is obtained for the heaviest alkanes (C28-36) it is expected that the general correlation fit these data best.
By utilizing a more complete specter of experimental data, the correlation will adapt to fit also the lighter alkanes better, possibly at the cost of the heavier alkanes fit.
To give each data-set equal influence, multiple regression was performed to find the $M_1$ and $M_2$ dependency of $A$ and $B$, respectively.
The resulting correlation did not differ significantly from \Eq{GenCorr} but generally gave less accurate predictions of the experimental data.

\section{Conclusions}
A review of published solubility data for binary n-alkane mixtures is presented.
Analysis of a total of 43 binary systems, from a total of 24 publications, revealed that there is a log-linear relationship between the solubility and the temperature, in the dilute range.
Data-set-specific linear regression was performed to obtain data-set-specific best-fit parameters for the solubility-temperature data, for solubilities below a mol-fraction of $0.1$, and it was seen that there is a clear linear relationship between the best-fit-parameters and the solvent and solute molar masses. 
Linear regression was thus employed to establish a general correlation between the solubility and the solvent and solute molar masses and the temperature.
Qualitative assessment shows that the developed correlation is successful at predicting the solubility trends seen in the experimental data, and reasonable predictions are obtained for the data-sets where no or only a few data-points exist below the mol-fraction of 0.1.
Furthermore, evidence is given that the correlation provides predictive power for multi-component mixtures by utilizing average solute and solvent molar masses as input.
More experiments are needed, however, for light solutes and a wider range of solvents in particular, to establish a more reliable correlation.
The developed general correlation should be used with caution outside its validity range and for mixtures for which it has not been tested.
\balance

\section*{Acknowledgements}
This study was financed by SINTEF.
The author is grateful for valuable feedback from colleagues and in particular Markus Hays Whitson (Whitson AS, Trondheim, Norway).




\bibliographystyle{unsrtnat}
\balance
\bibliography{References}

\begin{thebibliography}{44}
\providecommand{\natexlab}[1]{#1}
\providecommand{\url}[1]{\texttt{#1}}
\expandafter\ifx\csname urlstyle\endcsname\relax
  \providecommand{\doi}[1]{doi: #1}\else
  \providecommand{\doi}{doi: \begingroup \urlstyle{rm}\Url}\fi

\bibitem[Misra et~al.(1995)Misra, Baruah, and Singh]{Misra95}
S.~Misra, S.~Baruah, and K.~Singh.
\newblock Paraffin problems in crude oil production and transportation: A
  review.
\newblock \emph{SPE Production \& Facilities}, 10\penalty0 (1), 1995.
\newblock \doi{10.2118/28181-PA}.

\bibitem[Gluyas and Underhill(2003)]{Gluyas2003}
J.~G. Gluyas and J.~R. Underhill.
\newblock The staffa field, block 3/8b, uk north sea.
\newblock \emph{Geological Society, London, Memoirs}, 20\penalty0 (1):\penalty0
  327--333, 2003.
\newblock \doi{10.1144/GSL.MEM.2003.020.01.28}.

\bibitem[Coutinho et~al.(2006)Coutinho, Edmonds, Moorwood, Szczepanski, and
  Zhang]{Coutinho06}
J.~A.~P. Coutinho, B.~Edmonds, T.~Moorwood, R.~Szczepanski, and X.~Zhang.
\newblock Reliable wax predictions for flow assurance.
\newblock \emph{Energy \& Fuels}, 20\penalty0 (3):\penalty0 1081--1088, 2006.
\newblock \doi{10.1021/ef050082i}.

\bibitem[Heidariyan et~al.(2019)Heidariyan, Ehsani, Behbahani, and
  Mohammadi]{Heidariyan19}
H.~Heidariyan, M.~Ehsani, T.~J. Behbahani, and M.~Mohammadi.
\newblock Experimental investigation and thermodynamic modeling of wax
  precipitation in crude oil using the multi-solid model and {PC-SAFT EOS}.
\newblock \emph{Energy \& Fuels}, 33\penalty0 (10):\penalty0 9466--9479, 2019.
\newblock \doi{10.1021/acs.energyfuels.9b01445}.

\bibitem[Shahdi and Panacharoensawad(2019)]{Shahdi19}
A.~Shahdi and E.~Panacharoensawad.
\newblock {SP-Wax}: Solid-liquid equilibrium thermodynamic modeling software
  for paraffinic systems.
\newblock \emph{SoftwareX}, 9:\penalty0 145--153, 2019.
\newblock \doi{10.1016/j.softx.2019.01.015}.

\bibitem[Wang and Chen(2020)]{Wang20}
M.~Wang and C.C. Chen.
\newblock Predicting wax appearance temperature and precipitation profile of
  normal alkane systems: An explicit co-crystal model.
\newblock \emph{Fluid Phase Equilibria}, 509:\penalty0 112466, 2020.
\newblock \doi{10.1016/j.fluid.2020.112466}.

\bibitem[Ghasemi and Whitson(2021)]{Ghasemi21}
M.~Ghasemi and C.~H. Whitson.
\newblock {PVT} modeling of complex heavy oil mixtures.
\newblock \emph{Journal of Petroleum Science and Engineering}, 205:\penalty0
  108510, 2021.
\newblock \doi{10.1016/j.petrol.2021.108510}.

\bibitem[Singh et~al.(2000)Singh, Venkatesan, Fogler, and Nagarajan]{Singh2000}
P.~Singh, R.~Venkatesan, H.~S. Fogler, and N.~Nagarajan.
\newblock Formation and aging of incipient thin film wax-oil gels.
\newblock \emph{AIChE Journal}, 46\penalty0 (5):\penalty0 1059--1074, 2000.
\newblock \doi{10.1002/aic.690460517}.

\bibitem[Paso and Fogler(2004)]{Paso2004}
K.~G. Paso and H.~S. Fogler.
\newblock Bulk stabilization in wax deposition systems.
\newblock \emph{Energy and Fuels}, 18\penalty0 (4):\penalty0 1005--1013, 2004.
\newblock \doi{10.1021/ef034105+}.

\bibitem[Wu et~al.(2002)Wu, Wang, Shuler, Tang, Creek, Carlson, and
  Cheung]{Wu2004}
C.~H. Wu, K.~Wang, P.~J. Shuler, Y.~Tang, J.~L. Creek, R.~M. Carlson, and
  S.~Cheung.
\newblock Measurement of wax deposition in paraffin solutions.
\newblock \emph{AIChE Journal}, 48\penalty0 (9):\penalty0 2107--2110, 2002.
\newblock \doi{10.1002/aic.690480923}.

\bibitem[Johnsen et~al.(2011)Johnsen, Hanetho, Tetlie, Johansen, Einarsrud,
  Kaus, and Rone]{Johnsen11}
S.~G. Johnsen, S.~M. Hanetho, P.~Tetlie, S.~T. Johansen, M.~A. Einarsrud,
  I.~Kaus, and S.~C. Rone.
\newblock Paraffin wax deposition studies on coated and non-coated steel
  surfaces.
\newblock In \emph{Proceedings of Heat Exchanger Fouling and Cleaning IX -
  2011}, 2011.
\newblock \doi{10.13140/2.1.5085.6968}.

\bibitem[Seyer and Fordyce(1936)]{Seyer36}
W.~F. Seyer and R.~Fordyce.
\newblock The mutual solubilities of hydrocarbons. {I}. the freezing point
  curves of dotriacontane (dicetyl) in propane and butane.
\newblock \emph{Journal of the American Chemical Society}, 58\penalty0
  (10):\penalty0 2029--2031, 1936.
\newblock \doi{10.1021/ja01301a060}.

\bibitem[Godard(1937)]{Godard37}
H.~P. Godard.
\newblock The solubility of tetracosane in propane, butane, and pentane.
\newblock Master's thesis, University of British Columbia, 1937.
\newblock doi: 10.14288/1.0062154.

\bibitem[Campestrini and Stringari(2018)]{Campestrini18}
M.~Campestrini and P.~Stringari.
\newblock Solubilities of solid n-alkanes in methane: Data analysis and models
  assessment.
\newblock \emph{AIChE Journal}, 64\penalty0 (6):\penalty0 2219--2239, 2018.
\newblock \doi{10.1002/aic.16071}.

\bibitem[Tassin et~al.(2020)Tassin, {Rodríguez Reartes}, Zabaloy, and
  Cismondi]{Tassin20}
N.G. Tassin, S.B. {Rodríguez Reartes}, M.S. Zabaloy, and M.~Cismondi.
\newblock Modeling of solid-fluid equilibria of pure n-alkanes and binary
  methane + n-alkane systems through predictive correlations.
\newblock \emph{The Journal of Supercritical Fluids}, 166:\penalty0 105028,
  2020.
\newblock \doi{10.1016/j.supflu.2020.105028}.

\bibitem[Hoerr and Harwood(1951)]{Hoerr51}
C.~W. Hoerr and H.~J. Harwood.
\newblock Solubilities of high molecular weight aliphatic compounds in
  n-hexane.
\newblock \emph{The Journal of Organic Chemistry}, 16\penalty0 (5):\penalty0
  779--791, 1951.
\newblock \doi{10.1021/jo01145a020}.

\bibitem[Morawski et~al.(2005)Morawski, Coutinho, and Domanska]{Morawski05}
P.~Morawski, J.~A.~P. Coutinho, and U.~Domanska.
\newblock High pressure (solid+liquid) equilibria of n-alkane mixtures:
  experimental results, correlation and prediction.
\newblock \emph{Fluid Phase Equilibria}, 230:\penalty0 72 -- 80, 2005.
\newblock \doi{10.1016/j.fluid.2004.11.020}.

\bibitem[Dernini and De~Santis(1976)]{Dernini76}
S.~Dernini and R.~De~Santis.
\newblock Solubility of solid hexadecane and tetracosane in hexane.
\newblock \emph{The Canadian Journal of Chemical Engineering}, 54\penalty0
  (4):\penalty0 369--370, 1976.
\newblock \doi{10.1002/cjce.5450540421}.

\bibitem[Domanska(1990)]{SelData90}
U.~Domanska.
\newblock \emph{Selected Data on Mixtures: Series A}, chapter 10c. Solid-Liquid
  Equilibrium, pages 83--93.
\newblock Number~2 in International Data Series. Thermodynamics Research
  Center, 1990.

\bibitem[Chang et~al.(1983)Chang, Maurey, and Pummer]{Chang83}
S.~S. Chang, J.~R. Maurey, and W.~J. Pummer.
\newblock Solubilities of two n-alkanes in various solvents.
\newblock \emph{Journal of Chemical and Engineering Data}, 28\penalty0
  (2):\penalty0 187--189, 1983.
\newblock \doi{10.1021/je00032a017}.

\bibitem[Domanska et~al.(1987)Domanska, Hofman, and Rolinska]{Domanska87}
U.~Domanska, T.~Hofman, and J.~Rolinska.
\newblock Solubility and vapour pressures in saturated solutions of
  high-molecular-weight hydrocarbons.
\newblock \emph{Fluid Phase Equilibria}, 32\penalty0 (3):\penalty0 273 -- 293,
  1987.
\newblock \doi{10.1016/0378-3812(87)85059-8}.

\bibitem[Sadeghazad et~al.(2000)Sadeghazad, Christiansen, Sobhi, and
  Edalat]{Sadeghazad00}
A.~Sadeghazad, R.~L. Christiansen, G.~A. Sobhi, and M.~Edalat.
\newblock The prediction of cloud point temperature: In wax deposition.
\newblock Number SPE64519 in SPE Asia Pacific Oil and Gas Conference and
  Exhibition. SPE, 16-18 October 2000.
\newblock \doi{10.2118/64519-MS}.

\bibitem[Domanska(1987{\natexlab{a}})]{SelData87c}
U.~Domanska.
\newblock \emph{Selected Data on Mixtures: Series A}, chapter 10c. Solid-Liquid
  Equilibrium.
\newblock Number~4 in International Data Series. Thermodynamics Research
  Center, 1987{\natexlab{a}}.

\bibitem[Kniaz(1991)]{Kniaz91}
K.~Kniaz.
\newblock Solubility of n-docosane in n-hexane and cyclohexane.
\newblock \emph{Journal of Chemical and Engineering Data}, 36\penalty0
  (4):\penalty0 471--472, 1991.
\newblock \doi{10.1021/je00004a035}.

\bibitem[Fl\"oter et~al.(1997)Fl\"oter, Hollanders, de~Loos, and
  de~Swaan~Arons]{Floter97}
E.~Fl\"oter, B.~Hollanders, T.~W. de~Loos, and J.~de~Swaan~Arons.
\newblock The ternary system (n-heptane + docosane + tetracosane): The
  solubility of mixtures of docosane and tetracosane in heptane and data on
  solid-liquid and solid-solid equilibria in the binary subsystem (docosane +
  tetracosane).
\newblock \emph{Journal of Chemical and Engineering Data}, 42\penalty0
  (3):\penalty0 562--565, 1997.
\newblock \doi{10.1021/je960332c}.

\bibitem[Provost et~al.(1998)Provost, Chevallier, Bouroukba, Petitjean, and
  Dirand]{Provost98}
E.~Provost, V.~Chevallier, M.~Bouroukba, D.~Petitjean, and M.~Dirand.
\newblock Solubility of some n-alkanes ({C}23, {C}25, {C}26, {C}28) in heptane,
  methylcyclohexane, and toluene.
\newblock \emph{Journal of Chemical and Engineering Data}, 43\penalty0
  (5):\penalty0 745--749, 1998.
\newblock \doi{10.1021/je980027m}.

\bibitem[Brecevic and Garside(1993)]{Brecevic93}
L.~Brecevic and J.~Garside.
\newblock Solubilities of tetracosane in hydrocarbon solvents.
\newblock \emph{Journal of Chemical and Engineering Data}, 38\penalty0
  (4):\penalty0 598--601, 1993.
\newblock \doi{10.1021/je00012a032}.

\bibitem[Roberts et~al.(1994)Roberts, Rousseau, and Teja]{Roberts94}
K.~L. Roberts, R.~W. Rousseau, and A.~S. Teja.
\newblock Solubility of long-chain n-alkanes in heptane between 280 and 350
  {K}.
\newblock \emph{Journal of Chemical and Engineering Data}, 39\penalty0
  (4):\penalty0 793--795, 1994.
\newblock \doi{10.1021/je00016a035}.

\bibitem[Domanska(1987{\natexlab{b}})]{SelData87f}
U.~Domanska.
\newblock \emph{Selected Data on Mixtures: Series A}, chapter 10f. Solid-Liquid
  Equilibrium, pages 271--285.
\newblock Number~4 in International Data Series. Thermodynamics Research
  Center, 1987{\natexlab{b}}.

\bibitem[Ashbaugh et~al.(2002)Ashbaugh, Radulescu, Prud'homme, Schwahn,
  Richter, and Fetters]{Ashbaugh02}
H.~S. Ashbaugh, A.~Radulescu, R.~K. Prud'homme, D.~Schwahn, D.~Richter, and
  L.~J. Fetters.
\newblock Interaction of paraffin wax gels with random crystalline/amorphous
  hydrocarbon copolymers.
\newblock \emph{Macromolecules}, 35\penalty0 (18):\penalty0 7044--7053, 2002.
\newblock \doi{10.1021/ma0204047}.

\bibitem[Johnsen(2013)]{Johnsen12}
S.~G. Johnsen.
\newblock The solubility of n-tetracosane ({nC24}) in n-heptane ({nC7}) and
  n-decane ({nC10}).
\newblock \emph{Fuel}, 112:\penalty0 594--595, 2013.
\newblock \doi{10.1016/j.fuel.2012.03.015}.

\bibitem[Rakotosaona et~al.(2004)Rakotosaona, Bouroukba, Petitjean, Hubert,
  Mo{\"i}se, and Dirand]{Rakotosaona04}
R.~Rakotosaona, M.~Bouroukba, D.~Petitjean, N.~Hubert, J.~C. Mo{\"i}se, and
  M.~Dirand.
\newblock Binary phase diagram ({C}14+{C}25) and isopleth ({C}14+wax):
  comparison of solubility of n-pentacosane and multiparaffinic wax in
  n-tetradecane.
\newblock \emph{Fuel}, 83\penalty0 (7-8):\penalty0 851 -- 857, 2004.
\newblock \doi{10.1016/S0016-2361(03)00249-7}.

\bibitem[Madsen and Boistelle(1976)]{Madsen76}
H.E.~Lundager Madsen and R.~Boistelle.
\newblock Solubility of long-chain n-paraffins in pentane and heptane.
\newblock \emph{J. Chem. Soc., Faraday Trans. I}, 72:\penalty0 1078--1081,
  1976.
\newblock \doi{10.1039/F19767201078}.

\bibitem[Madsen and Boistelle(1979)]{Madsen79}
H.E.~Lundager Madsen and R.~Boistelle.
\newblock Solubility of octacosane and hexatriacontane in different n-alkane
  solvents.
\newblock \emph{J. Chem. Soc., Faraday Trans. I}, 75:\penalty0 1254--1258,
  1979.
\newblock \doi{10.1039/F19797501254}.

\bibitem[Seyer(1938)]{Seyer38}
W.~F. Seyer.
\newblock Mutual solubilities of hydrocarbons. ii. the freezing point curves of
  dotriacontane (dicetyl) in dodecane, decane, octane, hexane, cyclohexane and
  benzene.
\newblock \emph{Journal of the American Chemical Society}, 60\penalty0
  (4):\penalty0 827--830, 1938.
\newblock \doi{10.1021/ja01271a019}.

\bibitem[Hildebrand and Wachter(1949)]{Hildebrand49}
J.~H. Hildebrand and A.~Wachter.
\newblock The solubility of n-dotriacontane (dicetyl).
\newblock \emph{The Journal of Physical and Colloid Chemistry}, 53\penalty0
  (6):\penalty0 886--890, 1949.
\newblock \doi{10.1021/j150471a012}.

\bibitem[Jennings and Weispfennig(2005)]{Jennings05}
D.W. Jennings and K.~Weispfennig.
\newblock Experimental solubility data of various n-alkane waxes: effects of
  alkane chain length, alkane odd versus even carbon number structures, and
  solvent chemistry on solubility.
\newblock \emph{Fluid Phase Equilibria}, 227:\penalty0 27--35, 2005.
\newblock \doi{10.1016/j.fluid.2004.10.021}.

\bibitem[Weimer and Prausnitz(1965)]{Weimer65}
R.~F. Weimer and J.~M. Prausnitz.
\newblock Complex formation between carbon tetrachloride and aromatic
  hydrocarbons.
\newblock \emph{The Journal of Chemical Physics}, 42\penalty0 (10):\penalty0
  3643--3644, 1965.
\newblock \doi{10.1063/1.1695773}.

\bibitem[Choi and Mclaughlin(1983)]{Choi83}
P.~B. Choi and E.~Mclaughlin.
\newblock Effect of a phase transition on the solubility of a solid.
\newblock \emph{AIChE Journal}, 29\penalty0 (1):\penalty0 150--153, 1983.
\newblock \doi{10.1002/aic.690290121}.

\bibitem[Huber(1964)]{Huber64}
P.~J. Huber.
\newblock Robust estimation of a location parameter.
\newblock \emph{The Annals of Mathematical Statistics}, 35\penalty0
  (1):\penalty0 73--101, 1964.
\newblock \doi{10.1214/aoms/1177703732}.

\bibitem[Birkes and Dodge(1993)]{Birkes93}
D.~Birkes and Y.~Dodge.
\newblock \emph{Alternative Methods of Regression}.
\newblock Wiley-Interscience, 1993.
\newblock ISBN 9780471568810.

\bibitem[Dollhopf et~al.(1981)Dollhopf, Grossmann, and Leute]{Dollhopf81}
W.~Dollhopf, H.~P. Grossmann, and U.~Leute.
\newblock Some thermodynamic quantities of n-alkanes as a function of chain
  length.
\newblock \emph{Colloid \& Polymer Science}, 259:\penalty0 267--278, 1981.
\newblock \doi{10.1007/BF01381772}.

\bibitem[Laux et~al.(1999)Laux, Butz, Meyer, Matth\"ai, and Hildebrand]{Laux99}
H.~Laux, T.~Butz, G.~Meyer, M.~Matth\"ai, and G.~Hildebrand.
\newblock Thermodynamic functions of the solid-liquid transition of hydrocarbon
  mixtures.
\newblock \emph{Petroleum Science and Technology}, 17\penalty0 (9-10):\penalty0
  897--913, 1999.
\newblock \doi{10.1080/10916469908949755}.

\bibitem[Berne-Allen and Work(1938)]{Berne-Allen38}
A.~Berne-Allen and L.~T. Work.
\newblock Solubility of refined paraffin waxes in petroleum fractions.
\newblock \emph{Industrial and Engineering Chemistry}, 30\penalty0
  (7):\penalty0 806--812, 1938.
\newblock \doi{10.1021/ie50343a019}.

\end{thebibliography}
\balance


\newpage
\appendix
\renewcommand\thefigure{A.\arabic{figure}} 
\setcounter{figure}{0}
\onecolumn
\section*{Appendix: Detailed Comparison of Calculated and Experimental Data}
In the following figures, experimental solubility data-sets along with data-set-specific best-fit curves (for data below $0.1$ solute mol-fraction) and the general correlation given in \Eq{GenCorr} are shown for each data-set.
\newpage

\begin{figure*}[tb]
  \centering
  \subfloat[C8 in C6]{\includegraphics[width=0.5\linewidth]{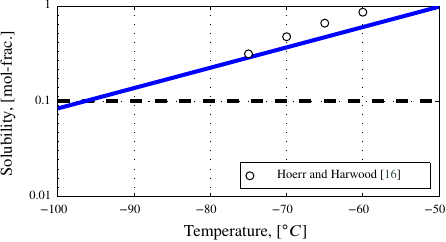}}
  \subfloat[C12 in C6]{\includegraphics[width=0.5\linewidth]{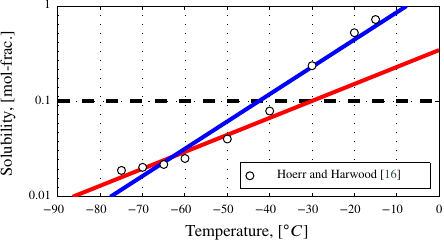}}\\
  \subfloat[C13 in C6\label{fig:C13C6}]{\includegraphics[width=0.5\linewidth]{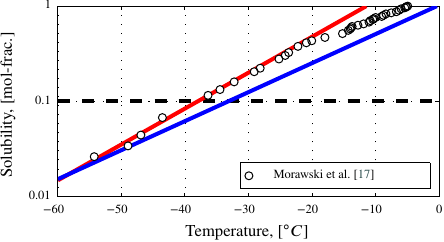}}
  \subfloat[C16 in C6\label{fig:C16C6}]{\includegraphics[width=0.5\linewidth]{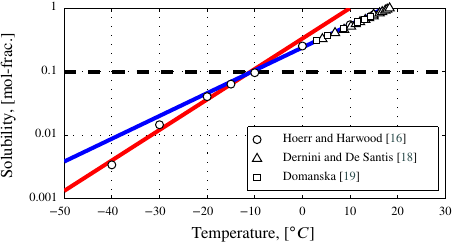}}\\
  \subfloat[C17 in C6]{\includegraphics[width=0.5\linewidth]{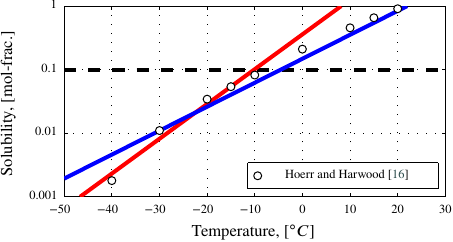}}
  \subfloat[C18 in C6\label{fig:C18C6}]{\includegraphics[width=0.5\linewidth]{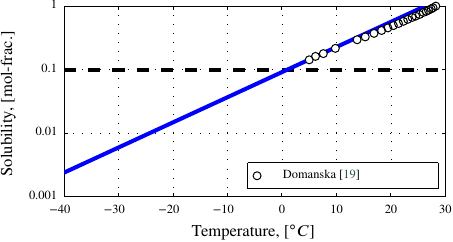}}\\
  \subfloat[C18 in C7\label{fig:C18C7}]{\includegraphics[width=0.5\linewidth]{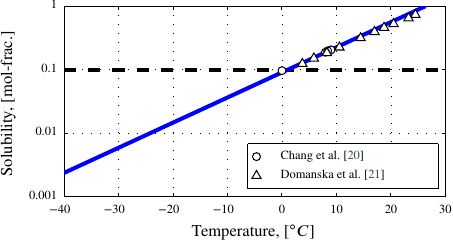}}
  \subfloat[C18 in C10\label{fig:C18C10}]{\includegraphics[width=0.5\linewidth]{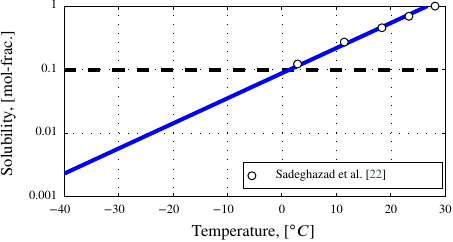}}\\
 \subfloat{\includegraphics[width=0.8\linewidth]{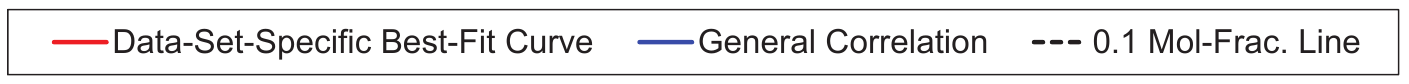}}
 \phantomcaption
\end{figure*}
\begin{figure*}[tb]
  \ContinuedFloat
  \addtocounter{subfigure}{-1} 
  \centering
  \subfloat[C19 in C7]{\includegraphics[width=0.5\linewidth]{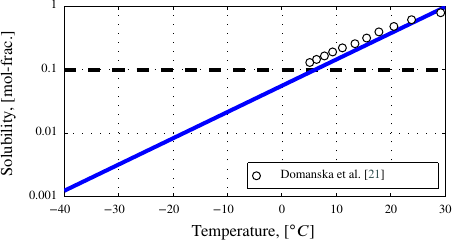}}
  \subfloat[C20 in C6\label{fig:C20C6}]{\includegraphics[width=0.5\linewidth]{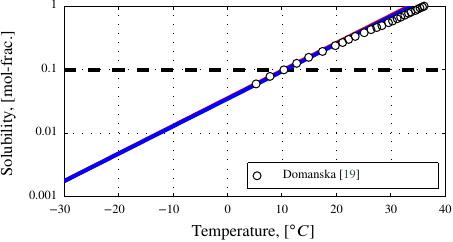}}\\
  \subfloat[C20 in C7]{\includegraphics[width=0.5\linewidth]{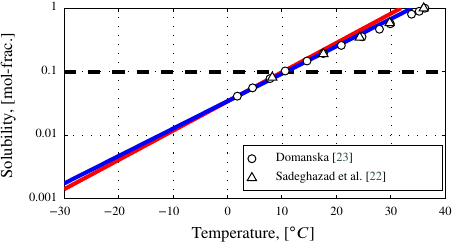}}
  \subfloat[C20 in C10]{\includegraphics[width=0.5\linewidth]{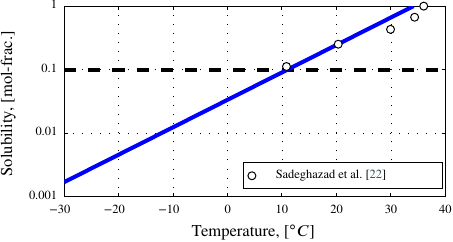}}\\
  \subfloat[C22 in C6]{\includegraphics[width=0.5\linewidth]{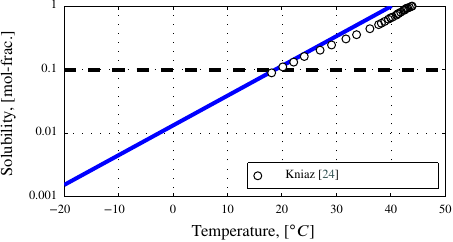}}
  \subfloat[C22 in C7]{\includegraphics[width=0.5\linewidth]{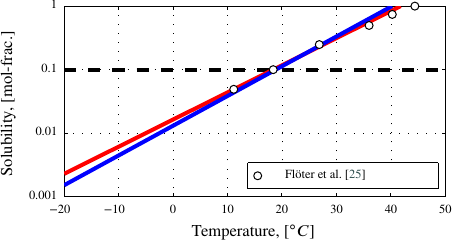}}\\
  \subfloat[C23 in C7]{\includegraphics[width=0.5\linewidth]{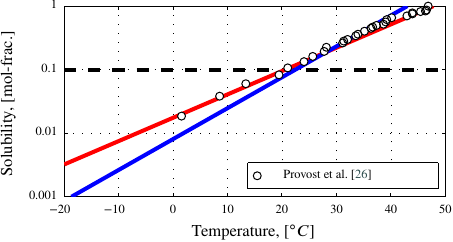}}\\
 \subfloat{\includegraphics[width=0.8\linewidth]{Legend}}
 \phantomcaption
\end{figure*}
\begin{figure*}
  \ContinuedFloat
  \addtocounter{subfigure}{-1} 
  \centering
  \subfloat[C24 in C3]{\includegraphics[width=0.5\linewidth]{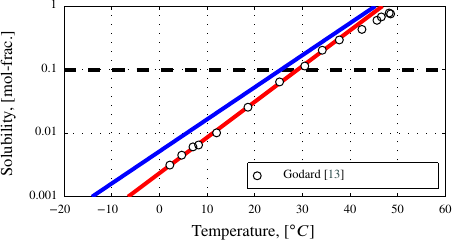}}
  \subfloat[C24 in C4]{\includegraphics[width=0.5\linewidth]{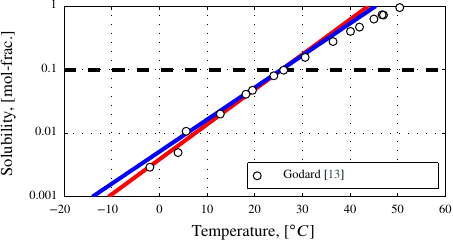}}\\
  \subfloat[C24 in C5]{\includegraphics[width=0.5\linewidth]{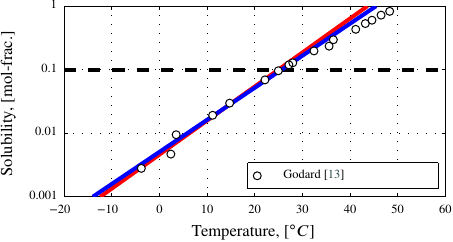}}
  \subfloat[C24 in C6]{\includegraphics[width=0.5\linewidth]{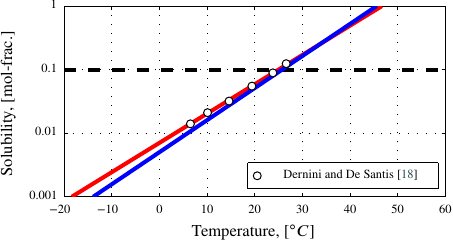}}\\
  \subfloat[C24 in C7]{\includegraphics[width=0.5\linewidth]{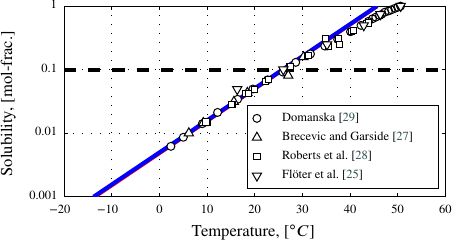}}
  \subfloat[C24 in C10]{\includegraphics[width=0.5\linewidth]{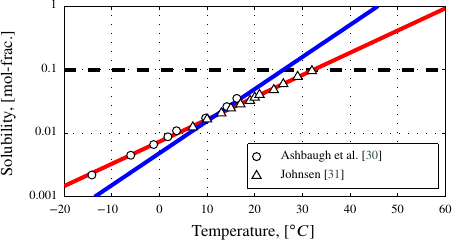}}\\
  \subfloat[C24 in C12]{\includegraphics[width=0.5\linewidth]{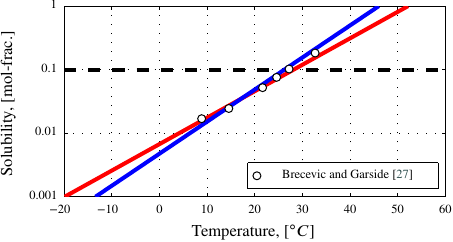}}\\
 \subfloat{\includegraphics[width=0.8\linewidth]{Legend}}
 \phantomcaption
\end{figure*}
\begin{figure*}[tb]
  \ContinuedFloat
  \addtocounter{subfigure}{-1} 
  \centering
  \subfloat[C25 in C7]{\includegraphics[width=0.5\linewidth]{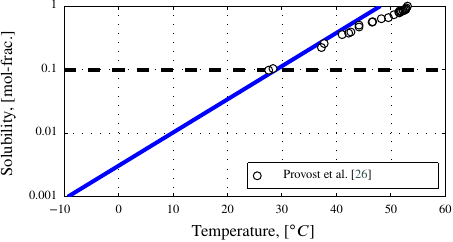}}
  \subfloat[C25 in C14]{\includegraphics[width=0.5\linewidth]{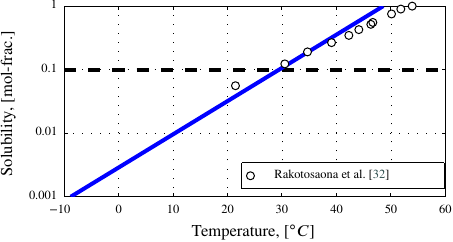}}\\
  \subfloat[C26 in C7]{\includegraphics[width=0.5\linewidth]{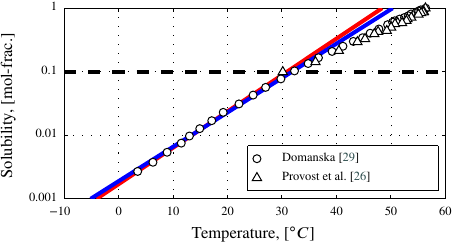}}\\
  \subfloat[C28 in C5]{\includegraphics[width=0.5\linewidth]{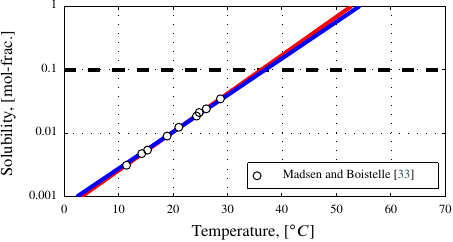}}
  \subfloat[C28 in C7]{\includegraphics[width=0.5\linewidth]{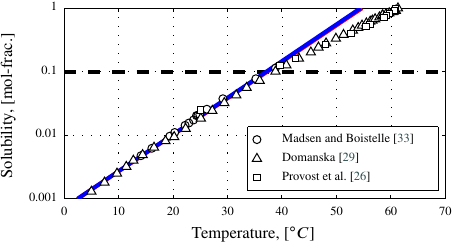}}\\
  \subfloat[C28 in C10]{\includegraphics[width=0.5\linewidth]{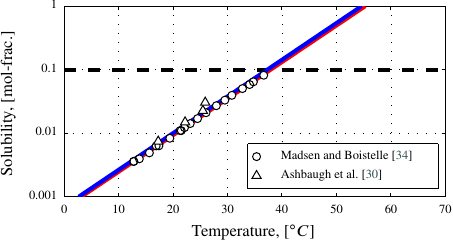}}
  \subfloat[C28 in C12]{\includegraphics[width=0.5\linewidth]{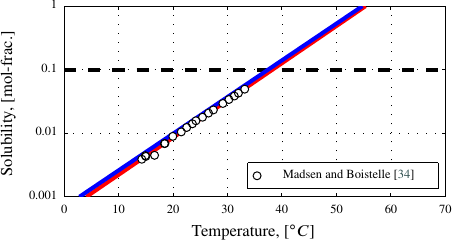}}\\
 \subfloat{\includegraphics[width=0.8\linewidth]{Legend}}
 \phantomcaption
\end{figure*}
\begin{figure*}[tb]
  \ContinuedFloat
  \addtocounter{subfigure}{-1} 
  \centering
  \subfloat[C32 in C3]{\includegraphics[width=0.5\linewidth]{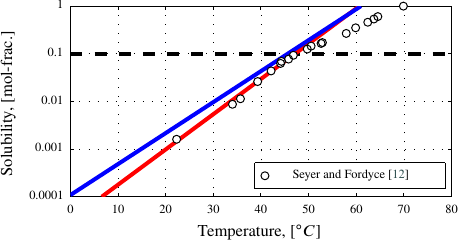}}
  \subfloat[C32 in C4]{\includegraphics[width=0.5\linewidth]{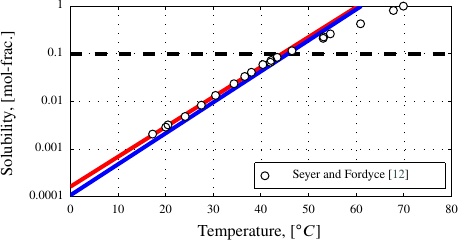}}\\
  \subfloat[C32 in C5]{\includegraphics[width=0.5\linewidth]{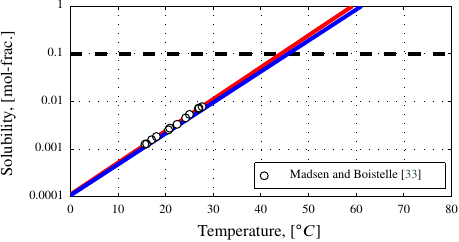}}
  \subfloat[C32 in C6]{\includegraphics[width=0.5\linewidth]{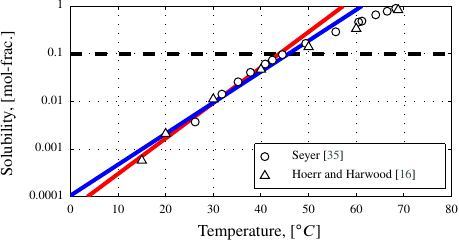}}\\
  \subfloat[C32 in C7]{\includegraphics[width=0.5\linewidth]{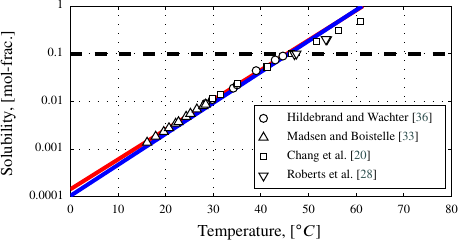}}
  \subfloat[C32 in C8]{\includegraphics[width=0.5\linewidth]{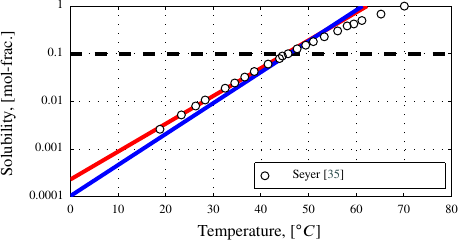}}\\
  \subfloat[C32 in C10]{\includegraphics[width=0.5\linewidth]{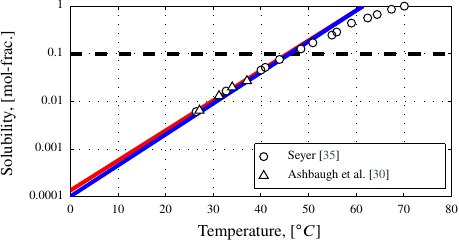}}
  \subfloat[C32 in C12]{\includegraphics[width=0.5\linewidth]{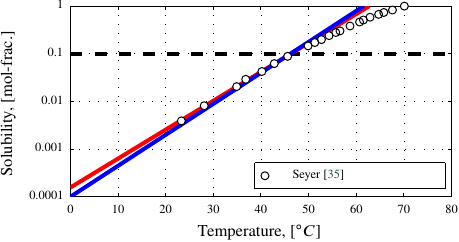}}\\
  \subfloat{\includegraphics[width=0.8\linewidth]{Legend}}
 \phantomcaption
\end{figure*}
\begin{figure*}[tb]
  \ContinuedFloat
  \addtocounter{subfigure}{-1} 
  \centering
  \subfloat[C36 in C5]{\includegraphics[width=0.5\linewidth]{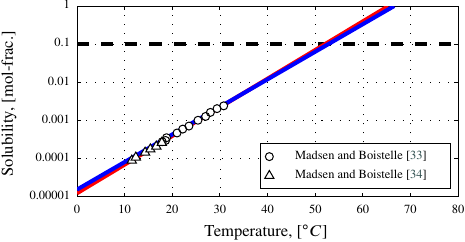}}
  \subfloat[C36 in C6]{\includegraphics[width=0.5\linewidth]{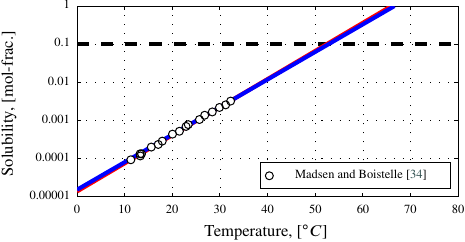}}\\
  \subfloat[C36 in C7]{\includegraphics[width=0.5\linewidth]{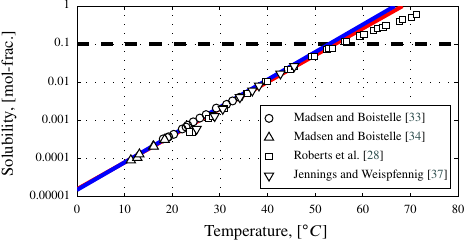}}
  \subfloat[C36 in C8]{\includegraphics[width=0.5\linewidth]{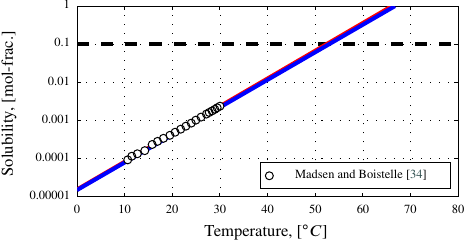}}\\
  \subfloat[C36 in C10\label{fig:C36C10}]{\includegraphics[width=0.5\linewidth]{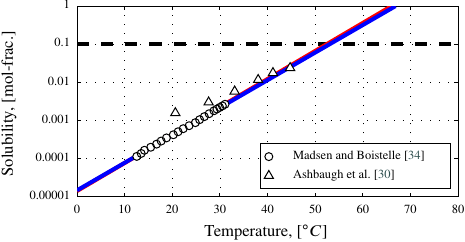}}
  \subfloat[C36 in C12]{\includegraphics[width=0.5\linewidth]{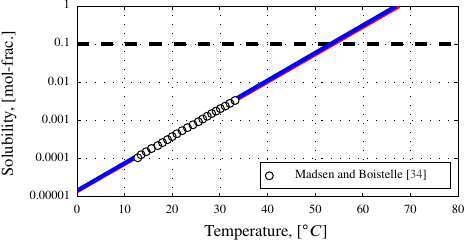}}\\
 \subfloat{\includegraphics[width=0.8\linewidth]{Legend}}
 \caption{Experimental solubility data-sets along with data-set-specific best-fit curves (for data below $0.1$ solute mol-fraction) and the general correlation given in \Eq{GenCorr}.}
\label{fig:SolubilityCurves}
\end{figure*}

\end{document}